\documentclass[aps, prd, twocolumn, superscriptaddress, preprintnumbers, nofootinbib]{revtex4}

\pdfoutput=1

\usepackage{amsmath, amssymb, amsfonts}
\usepackage{bm, bbm}
\usepackage{epsfig}

\newcommand{\Sec}[1]{section~\ref{sec:#1}}
\newcommand{\Secs}[1]{sections~\ref{sec:#1}}
\newcommand{\Secl}[1]{\label{sec:#1}}
\newcommand{\App}[1]{appendix~\ref{app:#1}}
\newcommand{\Appl}[1]{\label{app:#1}}
\newcommand{\Fig}[1]{figure~\ref{fig:#1}}
\newcommand{\Figl}[1]{\label{fig:#1}}
\newcommand{\Tab}[1]{table~\ref{tab:#1}}
\newcommand{\Tabl}[1]{\label{tab:#1}}

\newcommand{\nn}{\nonumber}
\newcommand{\eql}[1]{\label{eq:#1}}
\newcommand{\eq}[1]{(\ref{eq:#1})}

\newcommand{\e}{\mathrm{e}}
\newcommand{\I}{\mathrm{i}}

\newcommand{\DD}{\mathrm{D}}

\newcommand{\gsim}{\gtrsim}

\newcommand{\eqv}{\equiv}

\newcommand{\fr}[2]{\dfrac{#1}{#2}}

\newcommand{\dt}{\!\cdot\!}

\newcommand{\too}{\longrightarrow}

\newcommand{\sla}[1]{\setbox0=\hbox{$#1$}           % set a box for #1
   \dimen0=\wd0                                     % and get its size
   \setbox1=\hbox{/} \dimen1=\wd1                   % get size of /
   \ifdim\dimen0>\dimen1                            % if #1 is bigger than /
      \rlap{\hbox to \dimen0{\hfil/\hfil}}          % then center / in box
      #1                                            % and print #1
   \else                                            % if / is bigger than #1
      \rlap{\hbox to \dimen1{\hfil$#1$\hfil}}       % then center #1
      /                                             % and print /
   \fi} 

\newcommand{\Abs}[1]{\bigl\lvert #1 \bigr\rvert}

\newcommand{\al}{\alpha}
\newcommand{\be}{\beta}
\newcommand{\ga}{\gamma}
\newcommand{\Ga}{\Gamma}

\newcommand{\De}{\Delta}
\newcommand{\ep}{\epsilon}
\newcommand{\vep}{\varepsilon}

\newcommand{\la}{\lambda}
\newcommand{\La}{\Lambda}

\newcommand{\sg}{\sigma}

\newcommand{\tha}{\theta}
\newcommand{\Tha}{\Theta}

\newcommand{\cc}{\mathrm{c.c.}}

\newcommand{\cL}{\mathcal{L}}
\newcommand{\cM}{\mathcal{M}}
\newcommand{\cO}{\mathcal{O}}

\newcommand{\gu}[1]{\mathrm{U}(#1)}

\newcommand{\TeV}{\>\text{TeV}}

\newcommand{\iab}{\>\text{ab}^{-1}}

\newcommand{\LL}{\text{\tiny L}}
\newcommand{\RR}{\text{\tiny R}}

\newcommand{\pT}{p_\mathrm{T}}
\newcommand{\MET}{\sla{E}_\mathrm{T}}

\newcommand{\PD}{{\phantom{\dag}}}

\DeclareMathOperator{\Arccos}{Arccos}
\DeclareMathOperator{\sgn}{sgn}

\newcommand{\dsp}[1]{\displaystyle{#1}}

\newcommand{\Pboring}{P_\text{$\sla{\De}, \mathrm{S}$}}
\newcommand{\Puseless}{P_\text{$\mathrm{\De}, \sla{\mathrm{S}}$}}
\newcommand{\Pgood}{P_\text{$\De, \mathrm{S}$}}

\newcommand{\rest}{{}^{\!}\bigr|_0}

\usepackage{xcolor}
\definecolor{red}{rgb}{0.9, 0,0}

\begin{document}

\preprint{CERN-PH-TH/2013-181}
\preprint{FERMILAB-PUB-13-296-T}

\title{Measuring $CP$ Violation in $h \to \tau^+ \tau^-$ at Colliders}

\author{Roni Harnik}
\affiliation{Theoretical Physics Department, Fermilab, P.O.~Box 500,
  Batavia, IL 60510, USA}

\author{Adam Martin}
\affiliation{Department of Physics, University of Notre Dame, Notre
  Dame, IN 46556, USA} 
\affiliation{PH-TH Department, CERN, CH-1211 Geneva 23, Switzerland}

\author{Takemichi Okui} 
\affiliation{Department of Physics, Florida State University,
  Tallahassee, Florida 32306-4350, USA}

\author{Reinard Primulando}
\affiliation{Department of Physics and Astronomy, Johns Hopkins
  University, Baltimore, MD 21218, USA}

\author{Felix Yu}
\affiliation{Theoretical Physics Department, Fermilab, P.O.~Box 500,
  Batavia, IL 60510, USA}

\begin{abstract}
We investigate the LHC and Higgs Factory prospects for measuring the
$CP$ phase in the Higgs-$\tau$-$\tau$ coupling. Currently this phase
can be anywhere between $0^\circ$ ($CP$ even) and $90^\circ$ ($CP$
odd).  A new, ideal observable is identified from an analytic
calculation for the $\tau^\pm \to \rho^\pm\nu \to \pi^\pm \pi^0 \nu$
channel.  It is demonstrated to have promising sensitivity at the LHC
and superior sensitivity at the ILC compared to previous proposals.
Our observable requires the reconstruction of the internal
substructure of decaying taus but does not rely on measuring the
impact parameter of tau decays. It is the first proposal for such a
measurement at the LHC.  For the 14 TeV LHC, we estimate that about 1
ab$^{-1}$ data can discriminate $CP$-even versus $CP$-odd at the
$5\sigma$ level.  With 3 ab$^{-1}$, the $CP$ phase should be
measurable to an accuracy of $\sim 11^\circ$.  At an $e^+e^-$ Higgs
Factory, we project that a 250 GeV run with 1 ab$^{-1}$ luminosity can
measure the phase to $\sim 4.4^\circ$ accuracy.
\end{abstract}

\maketitle

%%%%%%%%%%%%%%%%%%%%%%%%%%%%%%%%%%%%%%%%%%%%%%%%%%%%%%%%%%%%%%%%%%%%%%%%%%%%%%%
%%%%%%%%%%%%%%%%%%%%%%%%%%%%%%%%%%%%%%%%%%%%%%%%%%%%%%%%%%%%%%%%%%%%%%%%%%%%%%%
%%%%%%%%%%%%%%%%%%%%%%%%%%%%%%%%%%%%%%%%%%%%%%%%%%%%%%%%%%%%%%%%%%%%%%%%%%%%%%%
\section{Introduction}

The discovery of the Higgs boson~\cite{Aad:2012tfa,
  Chatrchyan:2012ufa} has opened a new opportunity in the search for
physics beyond the SM\@.  The SM predicts all couplings of the Higgs
to SM particles completely, and a measured significant deviation of
Higgs couplings from the SM prediction will be a clear signal of new
physics.  The most straightforward tests at the moment are comparisons
of the Higgs production rates times branching ratios to the SM
prediction in a variety of final states. Thus far, such global fits
roughly agree with a SM Higgs~\cite{LHChiggs, Aaltonen:2013kxa}.

We can go further by testing the $CP$ properties of Higgs couplings.
This test has already been done for the coupling of the Higgs to
electroweak gauge bosons~\cite{Chatrchyan:2012jja, ATLAS:2013nma}.  In
the SM, the Higgs couples to the $Z$ boson as a scalar, $h Z_\mu
Z^\mu$. In general, a Higgs-like state could couple to $Z$ bosons as a
pseudoscalar, $h Z_{\mu \nu} \tilde Z^{\mu \nu} $, or with any linear
combination of scalar and pseudoscalar couplings, which would imply
$CP$ violation.  In the fully leptonic channel for $h \to Z Z^*$, the
azimuthal angle between the decay planes of the two $Z$ bosons is
sensitive to the $Z$ polarizations, which in turn is sensitive to the
$CP$ structure of the Higgs couplings~\cite{Bolognesi:2012mm}.
Current data disfavors a pure pseudoscalar coupling at 99.84\% = 3.3
$\sigma$ and 99.6\% (97.8\%) confidence level using CL$_\text{S}$
statistics at CMS~\cite{Chatrchyan:2012jja} and ATLAS MELA (ATLAS
BDT)~\cite{ATLAS:2013nma}, respectively.

In models where the SM is augmented by heavy new physics, this result
is unsurprising.  Of the two possible interactions mentioned above,
the scalar interaction is renormalizable, while the pseudoscalar
interaction arises from a dimension six operator.  The pseudoscalar
coupling is thus expected to be subdominant in Higgs decays and
corresponding $CP$ violating effects will be small.  While current
results favor a pure scalar coupling in the Higgs couplings to weak
gauge bosons, searches for $CP$ violation in fermionic decays of the
Higgs are still highly motivated. Such modified couplings can arise
from a different source which, in particular, can give a pseudoscalar
interaction comparable to a scalar, unlike the Higgs-$W$/$Z$
couplings.

In this paper we investigate how the $CP$ structure of the coupling of
the Higgs to tau leptons can be probed at present and future
colliders.  The Higgs coupling to any fermion generally consists of a
$CP$ even and a $CP$ odd term,
\begin{equation}
\mathcal{L}_{hff} \propto h \bar f (\cos \Delta + \I \ga_5 \sin \Delta) f \ .
\end{equation}
Measuring this $CP$ phase $\De$ requires knowledge of the spins of the
$f\bar{f}$ state.  Tau decays are complex enough to retain non-trivial
information about the direction of the tau spin, yet clean enough that
the spin information is not washed out by hadronization effects as it
is for $b$-quark decays~\cite{Grossman:2008qh}.  Since the Higgs
branching fraction to $\tau^+ \tau^-$ is substantial in the SM ($\sim
6.15\%$ for $m_h = 126$ GeV), the $\tau^+ \tau^-$ decay channel is the
best of a limited set of opportunities for $CP$ violation searches in
Higgs couplings to fermions.%
\footnote{For opportunities in other channels
  see~\cite{inprogress1, inprogress2}.}
In addition, a pseudoscalar-like coupling of the Higgs to taus can
conceivably compete with the small tau Yukawa coupling, and so $CP$
violating effects can be sizable. Currently, the only direct bound on
the Higgs-tau coupling is on the net signal strength in
$h\to\tau^+\tau^-$ channels: $\hat{\mu} = 1.1 \pm 0.4$~\cite{CMS:utj}.
In this paper, we will maintain $\hat{\mu} = 1$ and modify only $\De$,
so this constraint does not apply.

We focus on the specific tau decay channel $\tau^\pm\to \rho^\pm \nu$
with $\rho^\pm\to \pi^\pm \pi^0$.  This is the most common tau decay
sub-channel, with a branching fraction of $\sim 25\%$.  Moreover, the
angular distributions of the tau decay products and subsequent rho
decay products are correlated with the original direction of the tau
spin, as we will see in \Sec{heuristic}.  The relative azimuthal
orientation of the two hadronic taus, $\Theta$, which we will define
precisely in \Sec{Theta}, contains information about the $CP$
properties of the Higgs coupling to taus.  In particular, the $CP$
phase $\Delta$ in the Higgs couplings may be read off directly from
the $\Theta$ distribution.  The differential cross section is shown
analytically in~\Secs{EandB} and C to have the form $c - A \cos
(\Theta - 2\Delta)$, and $\Delta$ may be measured by finding the
minimum of the distribution (as exemplified in \Fig{True_varyDelta}).
The dominant background for $h \to \tau \tau$ events at the LHC is $Z
\to \tau \tau$, which produces a flat $\Theta$ distribution.

The ability to distinguish scalar versus pseudoscalar Higgs couplings
in the tau channel has been discussed in~\cite{Dell'Aquila:1988fe,
  Dell'Aquila:1988rx, Grzadkowski:1995rx, Bower:2002zx, Worek:2003zp,
  Desch:2003mw, Desch:2003rw, Berge:2008dr, Berge:2008wi,
  Berge:2011ij, Berge:2012wm}.  Our work quantitatively improves on
these results: our $\Theta$ variable is demonstrably more sensitive to
the $CP$ phase of the Higgs coupling to taus compared to earlier
proposed observables, and our simulation results for the ILC indicate
a corresponding increase in sensitivity compared to earlier results.
This work is also a qualitative step forward in that we propose a
strategy to do this measurement at the LHC.  Previous studies relied
on resolving a displaced vertex in $\tau$ decays which is challenging.
We show that our observable retains sensitivity without this.

It should be stressed that in order to reconstruct the angle $\Theta$,
full knowledge of all four-momenta components in the event is needed,
including those of the two neutrinos.  We will discuss the challenges
that this presents and how they may be addressed.  In the context of a
Higgs factory (ILC), $h \rightarrow \tau^+\tau^- \rightarrow \rho^+
\rho^- \nu \bar{\nu}$ events may be fully reconstructed up to a
two-fold ambiguity.  Furthermore, a favorable signal to background
ratio makes our measurement straightforward.  At a hadron collider,
however, some approximations are needed for the neutrino four-momenta.
Employing the collinear approximation~\cite{Ellis:1987xu}, we show
that the amplitude of the angular structure in $\Theta$ is only
reduced by an order one factor for $h \to \tau \tau$ signal events.
The challenge for the LHC is thus to increase the signal to background
ratio as much as possible in order to produce a statistically
significant result.  In addition, an improvement over the collinear
approximation would make a positive impact on the resulting
sensitivity to $\Delta$.

Our net result is that, using the $\Theta$ variable, a measurement of
$\Delta$ with an accuracy of $4.4^\circ$ is possible for a $\sqrt{s} =
250$ GeV $e^+ e^-$ collider, assuming 1 ab$^{-1}$ of luminosity
(without incorporating detector effects, which are expected to be
negligibly small). This number should be compared with the result of
Ref.~\cite{Desch:2003rw}, which quotes an accuracy of measuring
$\Delta$ to $6^\circ$ using the same amount of luminosity but for
$\sqrt{s} = 350$ GeV and $m_h = 120$ GeV\@. We also provide the first
estimates for sensitivity to $\Delta$ at the LHC\@.  Without
incorporating detector effects or pileup, we find an ideal measurement
of $\Delta$ to an accuracy of $11.5^\circ$ is possible with 3
ab$^{-1}$ of $\sqrt{s} = 14$ TeV LHC data for a $\tau$-tagging
efficiency of 50\%.  Improving the efficiency from 50\% to 70\% could
lead to an accuracy of $8.0^\circ$ using the same LHC luminosity.

This paper is organized as follows.  In \Sec{CPV_coupling} we add $CP$
violation to the Higgs coupling to tau leptons. In \Sec{observable} we
introduce our observable, first in a heuristic analysis that follows
every step of the decay, and then rigorously, using the analytic form
of the full $1\to6$ differential cross section. We present the results
of our collider analyses in \Sec{studies}.  We first present the
relevant distributions using Monte Carlo truth information, then
reevaluate in a Higgs factory setup, where a twofold ambiguity needs
to be considered, and finally consider an LHC setting using the
collinear approximation.  We conclude in \Sec{conclusion}.  A
weakly-coupled renormalizable model giving rise to $CP$ violation in
the Higgs coupling to taus is presented in \App{UVcompletion}.

%%%%%%%%%%%%%%%%%%%%%%%%%%%%%%%%%%%%%%%%%%%%%%%%%%%%%%%%%%%%%%%%%%%%%%%%%%%%%%%
%%%%%%%%%%%%%%%%%%%%%%%%%%%%%%%%%%%%%%%%%%%%%%%%%%%%%%%%%%%%%%%%%%%%%%%%%%%%%%%
%%%%%%%%%%%%%%%%%%%%%%%%%%%%%%%%%%%%%%%%%%%%%%%%%%%%%%%%%%%%%%%%%%%%%%%%%%%%%%%
\section{A $CP$-violating $h\tau\bar{\tau}$ coupling}
\Secl{CPV_coupling}
%%%%%%%%%%%%%%%%%%%%%%%%%%%%%%%%%%%%%%%%%%%%%%%%%%%%%%%%%%%%%%%%%%%%%%%%%%%%%%%
%%%%%%%%%%%%%%%%%%%%%%%%%%%%%%%%%%%%%%%%%%%%%%%%%%%%%%%%%%%%%%%%%%%%%%%%%%%%%%%
%%%%%%%%%%%%%%%%%%%%%%%%%%%%%%%%%%%%%%%%%%%%%%%%%%%%%%%%%%%%%%%%%%%%%%%%%%%%%%%
In our study of the $CP$ nature of $h \to \tau^+ \tau^-$, we use the
following phenomenological Lagrangian:
\begin{align}
\cL_\text{pheno} 
&\supset
-m _\tau \, \bar{\tau} \tau 
-\fr{y_\tau}{\sqrt2} \, h \bar{\tau} (\cos\De + \I \ga_5 \sin\De) \tau 
\nn\\%%
&= 
-m_\tau \, \bar{\tau} \tau 
-\fr{y_\tau}{\sqrt2} \, h \bigl( \tau^\dag_\LL (\cos\De + \I \sin\De) \tau_\RR 
\nn\\
&\hspace{20ex}
+\cc \bigr)
\,,\eql{Lpheno}
\end{align}
where $\tau$ and $h$ are the physical tau lepton and Higgs boson in
the mass basis, respectively, $y_\tau$ is a real parameter
parametrizing the magnitude of the $h \tau \bar{\tau}$ coupling, and,
most importantly, $\De \in (-\pi/2, \pi/2]$ is an angle describing the
  $CP$ nature of the $h \tau \bar{\tau}$ coupling.%
\footnote{The angle $\Delta$ can, in fact, take the full range of
  $(-\pi,\pi]$. However our technique is not sensitive to a
multiplication of the tau Yukawa by $-1$ and so it is sufficient to
consider half of this range. Resolving this ambiguity would require
measuring the interference of Higgs with background, which is a tiny
effect.}
The $CP$-even and $CP$-odd cases correspond to $\De = 0$ and $\De =
\pi/2$, respectively, while $\De = \pm \pi/4$ describe maximally
$CP$-violating cases.  The SM corresponds to a special case, $y_\tau =
y_\tau^\text{SM} \eqv m_\tau / v$ with $\De=0$.  We will refer to
``$\cos\De + \I \sin\De$'' as a ``$CP$-violating $h\tau\bar{\tau}$
coupling'', even though it includes the $CP$-conserving limits of $\De
= 0$ and $\pi/2$.  In this work, we focus on the effects of $\De$, so
we will take $y_\tau = y_\tau^\text{SM}$ while treating $\De$ as a
free parameter.

The simplest fully gauge-invariant operator that results in the
$CP$-violating $h \tau \bar{\tau}$ coupling~\eq{Lpheno} upon
electroweak symmetry breaking is given by
\begin{align}
\cL_\text{eff} 
&\supset 
-\Bigl( \al + \be \fr{H^\dag H}{\La^2} \Bigr) H \ell_{3\LL}^\dag \tau_\RR^\PD
+\cc
\,,\eql{Leff}
\end{align}
where $\al$ and $\be$ are \emph{complex} dimensionless parameters, and
$\La$ is a mass scale taken to be real and positive without loss of
generality.  To relate the parameters of $\cL_\text{pheno}$ and
$\cL_\text{eff}$, we substitute $H = \bigl( 0,\, v + h / \sqrt2
\bigr)^\mathrm{\! T}$ in~\eq{Leff}, which yields
\begin{align}
\cL_\text{eff} 
&\supset
-\Bigl( \al + \be \fr{v^2}{\La^2} \Bigr) v \tau_\LL^\dag \tau_\RR^\PD
-\Bigl( \al + 3\be \fr{v^2}{\La^2} \Bigr) \fr{h}{\sqrt2} \tau_\LL^\dag \tau_\RR^\PD
\nn\\
&\hspace{2.7ex}
+\cc 
\,,
\end{align}
from which we identify
\begin{align}
\al + \be \fr{v^2}{\La^2} 
=
y_\tau^\text{SM}
>0
\,, \eql{Lysm}
\end{align}
and we have taken $y_\tau^\text{SM}$ to be real and positive (hence
$m_\tau \equiv y_\tau^\text{SM} v$ is real and positive) without loss
of generality after suitable redefinition of the phase of $\tau_\RR$.
With this phase convention, the $h \tau \bar{\tau}$ coupling
in~\eq{Lpheno} is generally complex:
\begin{align}
y_\tau (\cos\De + \I \sin\De)
&=
\al + 3 \be \fr{v^2}{\La^2} 
\nn\\
&=
y_\tau^\text{SM} + 2 \be \fr{v^2}{\La^2}
\,.
\end{align}
Since $y_\tau^\text{SM} \sim 10^{-2}$, new physics at the TeV scale 
($\La \sim 10 v$) with $\cO(1)$ couplings ($|\beta| \sim 1$) 
can give rise to $\De$ anywhere in the full range $(-\pi/2, \pi/2]$.%
\footnote{An ``existence proof'' of such new physics in terms of a
  weakly-coupled renormalizable theory is given in
  \App{UVcompletion}.}
This is in stark contrast to the case of a $CP$-odd/violating Higgs
coupling to $Z$ bosons, where TeV-scale new physics is expected to
give only small corrections to the SM $CP$-even coupling.

%%%%%%%%%%%%%%%%%%%%%%%%%%%%%%%%%%%%%%%%%%%%%%%%%%%%%%%%%%%%%%%%%%%%%%%%%%%%%%%
%%%%%%%%%%%%%%%%%%%%%%%%%%%%%%%%%%%%%%%%%%%%%%%%%%%%%%%%%%%%%%%%%%%%%%%%%%%%%%%
%%%%%%%%%%%%%%%%%%%%%%%%%%%%%%%%%%%%%%%%%%%%%%%%%%%%%%%%%%%%%%%%%%%%%%%%%%%%%%%
\section{The observable}
\Secl{observable}
%%%%%%%%%%%%%%%%%%%%%%%%%%%%%%%%%%%%%%%%%%%%%%%%%%%%%%%%%%%%%%%%%%%%%%%%%%%%%%%
%%%%%%%%%%%%%%%%%%%%%%%%%%%%%%%%%%%%%%%%%%%%%%%%%%%%%%%%%%%%%%%%%%%%%%%%%%%%%%%
%%%%%%%%%%%%%%%%%%%%%%%%%%%%%%%%%%%%%%%%%%%%%%%%%%%%%%%%%%%%%%%%%%%%%%%%%%%%%%%
To probe the $CP$-violating $h \tau \bar{\tau}$ coupling
in~\eq{Lpheno}, we will study the following decay process:
\begin{align}
h 
&\>\too\> 
\tau^- \tau^+
\nn\\
&\>\too\> 
\rho^- \nu_\tau 
\> 
\rho^+ \bar{\nu}_\tau
\nn\\
&\>\too\> 
\pi^- \pi^0 
\> 
\nu_\tau 
\> 
\pi^+ \pi^0 
\> 
\bar{\nu}_\tau
\,.\eql{theprocess}
\end{align}
There are several good reasons to choose this decay chain.  First, to
minimize the loss of kinematic information due to neutrinos, we want
both $\tau^-$ and $\tau^+$ to decay hadronically.  Second, of the
hadronic decay modes, we choose $\tau \to \rho \nu$, since the
subsequent decay, $\rho^\pm \to \pi^\pm \pi^0$, can be reconstructed
at a collider.  Third, $\tau \to \rho \nu$ has the largest branching
fraction of any individual tau decay mode, $\sim 25$\%, and the
following step, $\rho \to \pi\pi$, occurs with a nearly 100\%
probability.  Finally, the $\rho$ width is sufficiently narrow that it
is well justified to consider it on-shell, which makes the process
in~\eq{theprocess} an analytically tractable sequence of 2-body
decays.

We begin with a heuristic look at the process in~\eq{theprocess} to
develop a rough idea of how it can probe the $CP$-violating $h \tau
\tau$ coupling~\eq{Lpheno}.  In particular, the highlights of
qualitative points to be made in sections~\ref{sec:h_to_tautau},
A$\,$2 and A$\,$3 are:
\begin{itemize}
\item[1:]{Measuring $\tau$ helicities cannot determine the $CP$ phase,
  but the $\tau$ polarizations in directions \emph{perpendicular} to
  the $\tau$ momenta can.}
\item[2:]{In the tau rest frame the $\rho$ is predominantly
  longitudinal and is polarized roughly in the direction of the $\tau$
  polarization.}
\item[3:]{The difference between the charged and neutral pion
  3-momenta, $\vec{p}_{\pi^\pm\!} - \vec{p}_{\pi^0}$, is roughly
  parallel to the respective $\rho^\pm$ polarization.}
\end{itemize}
Therefore, the $CP$ nature of $h \to \tau\tau$ must be encoded in the
orientation of ``$\vec{p}_{\pi^\pm\!} - \vec{p}_{\pi^0}$'' in the
plane perpendicular to the $\tau^\pm$ momenta in the Higgs rest frame.
A precise form of ``$\vec{p}_{\pi^\pm} - \vec{p}_{\pi^0}$'' as well as
the best observable to measure the $CP$ phase $\De$ will be identified
in sections~\ref{sec:EandB} and C by analytically computing the full
matrix element for the sequence of two-body decays in
process~\eq{theprocess}.

%%%%%%%%%%%%%%%%%%%%%%%%%%%%%%%%%%%%%%%%%%%%%%%%%%%%%%%%%%%%%%%%%%%%%%%%%%%%%%%
%%%%%%%%%%%%%%%%%%%%%%%%%%%%%%%%%%%%%%%%%%%%%%%%%%%%%%%%%%%%%%%%%%%%%%%%%%%%%%%
\subsection{A heuristic analysis}
\Secl{heuristic}
%%%%%%%%%%%%%%%%%%%%%%%%%%%%%%%%%%%%%%%%%%%%%%%%%%%%%%%%%%%%%%%%%%%%%%%%%%%%%%%
%%%%%%%%%%%%%%%%%%%%%%%%%%%%%%%%%%%%%%%%%%%%%%%%%%%%%%%%%%%%%%%%%%%%%%%%%%%%%%%

%%%%%%%%%%%%%%%%%%%%%%%%%%%%%%%%%%%%%%%%%%%%%%%%%%%%%%%%%%%%%%%%%%%%%%%%%%%%%%%
\subsubsection{$h \to \tau^- \, \tau^+$}
\Secl{h_to_tautau}
%%%%%%%%%%%%%%%%%%%%%%%%%%%%%%%%%%%%%%%%%%%%%%%%%%%%%%%%%%%%%%%%%%%%%%%%%%%%%%%
The most general form of the amplitude for the decay $h \to \tau^- \,
\tau^+$ is given by
\begin{align}
\cM_{h \to \tau\tau}
\propto
\sum_{s,s'} \chi_{s,s'} \,
\bar{u}_{\tau^-}^{s} \,  
(\cos\De + \I \ga_5 \sin\De) \,
v_{\tau^+}^{s'}  
\,,
\eql{M:h_to_tautau}
\end{align}
where $\chi_{s,s'}$ is the probability amplitude of $\tau^-$ and
$\tau^+$ having helicities $s/2$ and $s'/2$, respectively.  Lorentz
invariance dictates that the proportionality factor omitted
in~\eq{M:h_to_tautau} has no momentum dependence.

In the Higgs rest frame, the amplitude~\eq{M:h_to_tautau} takes the
form
\begin{align}
\cM_{h \to \tau\tau}
\propto\>
|\vec{p}_{\tau^-\!}| \, \chi^1_0 \cos\De
-
\I E_{\tau^-} \chi^0_0 \sin\De
\,,\eql{M:h_to_tautau:higgs_frame}
\end{align}
where $\vec{p}_{\tau^-}$ and $E_{\tau^-}$ are the $\tau^-$ momentum
and energy in this frame, while $\chi^j_m$ is the linear combination
of $\chi_{s,s'}$ with angular momentum $(j,m)$. In particular,
\begin{equation}
\chi^1_0 = \fr{\chi_{1,1} + \chi_{-1,-1}}{\sqrt2}
\,,\quad
\chi^0_0 = \fr{\chi_{1,1} - \chi_{-1,-1}}{\sqrt2}
\,.\eql{chi10_chi00}
\end{equation}
The amplitude in~\eq{M:h_to_tautau:higgs_frame} shows that the
$CP$-even contribution ($\propto \cos\De$) is a spin triplet in a
$p$-wave, while the $CP$-odd contribution ($\propto \sin\De$) is a
spin singlet in an $s$-wave.  This can be understood as a consequence
of angular momentum conservation and Fermi statistics, with the
additional fact that a fermion--anti-fermion pair has an odd intrinsic
parity.

To measure $\De$, it is necessary to keep the $\tau^-\tau^+$ pair in
the above superpositions of $\chi_{1,1}$ and $\chi_{-1,-1}$, without
projecting the polarizations onto the helicity eigenstates.
From~\eq{M:h_to_tautau:higgs_frame} and~\eq{chi10_chi00}, we see that
the coefficients of $\chi_{1,1}$ and $\chi_{-1,-1}$ are the complex
conjugates of each other, which implies that, regardless of $\De$, the
probability for both $\tau^-$ and $\tau^+$ to be right-handed is
always equal to that for both to be left-handed.  Therefore, to
distinguish the two linear combinations in~\eq{chi10_chi00}, we must
measure the polarizations in the directions \emph{perpendicular} to
the momenta, as mentioned in item 1 above.

%%%%%%%%%%%%%%%%%%%%%%%%%%%%%%%%%%%%%%%%%%%%%%%%%%%%%%%%%%%%%%%%%%%%%%%%%%%%%%%
\subsubsection{$\tau^- \to \rho^- \, \nu_\tau$}
%%%%%%%%%%%%%%%%%%%%%%%%%%%%%%%%%%%%%%%%%%%%%%%%%%%%%%%%%%%%%%%%%%%%%%%%%%%%%%%
Assuming the SM weak interactions for the $\tau$ and
$\nu_\tau$, the most general form of the amplitude for $\tau^- \to
\rho^- \, \nu_\tau$ is given by
\begin{align}
\mathcal{M}_{\tau \to \rho\nu}
\propto\> 
(\ep_{\rho^-\!}^*)_{\!\mu}^\PD \; 
\bar{u}_{\nu_\tau} \ga^\mu P_\LL \, u_{\tau^-} 
\,,\eql{M:tau_to_rho-nu}
\end{align}
with $P_\LL \equiv (1 - \ga_5) / 2$.  Again, Lorentz invariance
dictates that the proportionality factor omitted
in~\eq{M:tau_to_rho-nu} has no momentum dependence.

In the $\tau^-$ rest frame, the amplitude~\eq{M:tau_to_rho-nu} has the
form
\begin{align}
\cM_{\tau \to \rho\nu}
\propto\>
\ep_{\rho^-}^* \dt 
\Bigl(
\vep_{-1} \, \sin\frac{\theta}{2}
-\vep_0 \, \frac{m_\tau}{\sqrt2 \, m_\rho}  \,\cos\frac{\theta}{2}
\Bigr)
\,,\eql{M:tau_to_rho-nu:tau_frame}
\end{align}
where $\tha \in [0, \pi]$ is the angle between the $\rho^-$ momentum
and the $\tau^-$ polarization in this frame, and $\vep_{-1}^\mu$,
$\vep_0^\mu$, and $\vep_1^\mu$ are the polarization vectors for the
left-handed, longitudinal, and right-handed polarizations of the
$\rho^-$, respectively.  Since $m_\tau^2 / (2 m_\rho^2) \sim 3$, the
amplitude~\eq{M:tau_to_rho-nu:tau_frame} is dominated by the second
term, roughly speaking.  Thus, we are led to the picture described in
the item 2 above, namely, the $\rho^-$ is predominantly longitudinal
($\ep_{\rho^-} \sim \vep_0$) and mostly emitted in the direction of
the $\tau^-$ polarization ($\tha \sim 0$).

%%%%%%%%%%%%%%%%%%%%%%%%%%%%%%%%%%%%%%%%%%%%%%%%%%%%%%%%%%%%%%%%%%%%%%%%%%%%%%%
\subsubsection{$\rho^- \to \pi^- \, \pi^0$}
%%%%%%%%%%%%%%%%%%%%%%%%%%%%%%%%%%%%%%%%%%%%%%%%%%%%%%%%%%%%%%%%%%%%%%%%%%%%%%%
The most general form of the amplitude for $\rho^- \to \pi^- \, \pi^0$
is given by
\begin{align}
\mathcal{M}_{\rho \to \pi\pi}
\propto\>
\ep_{\rho^-} \dt (p_{\pi^-} - p_{\pi^0\!})
\,.\eql{M:rho_to_pi-pi}
\end{align}
The other linear combination, $p_{\pi^-\!} + p_{\pi^0}$, cannot appear
here because $\ep_{\rho^-\!} \cdot\! (p_{\pi^-\!} + p_{\pi^0}) =
\ep_{\rho^-\!} \cdot p_{\rho^-} = 0$.  Again, the proportionality
factor omitted in~\eq{M:rho_to_pi-pi} cannot have any momentum
dependence by Lorentz invariance.

Boosting the longitudinal $\rho^-$ to its rest frame, and neglecting
the $\pi^\pm$-$\pi^0$ mass difference, the
amplitude~\eq{M:rho_to_pi-pi} takes the form
\begin{align}
\mathcal{M}_{\rho \to \pi\pi}
\propto\>
|\vec{p}_{\pi^-\!} - \vec{p}_{\pi^0\!}| \cos\psi
\,,\eql{M:rho_to_pi-pi:rho_frame}
\end{align}
where $\psi$ is the angle between the original $\rho^-$ polarization
and the vector $\vec{p}_{\pi^-\!} - \vec{p}_{\pi^0}$ in the rest
frame.  Therefore, the momentum difference, $\vec{p}_{\pi^-\!}  -
\vec{p}_{\pi^0}$, is roughly (anti-)parallel ($\psi \sim 0$ or $\pi$)
to the original $\rho^-$ polarization, as we described in the item 3
above.

%%%%%%%%%%%%%%%%%%%%%%%%%%%%%%%%%%%%%%%%%%%%%%%%%%%%%%%%%%%%%%%%%%%%%%%%%%%%%%%
%%%%%%%%%%%%%%%%%%%%%%%%%%%%%%%%%%%%%%%%%%%%%%%%%%%%%%%%%%%%%%%%%%%%%%%%%%%%%%%
\subsection{The ``electric'' and ``magnetic'' variables}
\Secl{EandB}
%%%%%%%%%%%%%%%%%%%%%%%%%%%%%%%%%%%%%%%%%%%%%%%%%%%%%%%%%%%%%%%%%%%%%%%%%%%%%%%
%%%%%%%%%%%%%%%%%%%%%%%%%%%%%%%%%%%%%%%%%%%%%%%%%%%%%%%%%%%%%%%%%%%%%%%%%%%%%%%
We now analytically compute the full matrix element for the
process~\eq{theprocess} to identify the observable that is most
sensitive to the $CP$ phase $\De$.  Combining the
amplitudes~\eq{M:h_to_tautau},~\eq{M:tau_to_rho-nu}
and~\eq{M:rho_to_pi-pi}, the full amplitude for the
process~\eq{theprocess} at tree level is given by
\begin{align}
\cM_\text{full}
\propto\>\>&
\bar{u}_{\nu^-} 
(\sla{p}_{\pi^-\!} - \sla{p}_{\pi^{0-\!}}) \,
P_\LL \,
(\sla{p}_{\tau^-\!} + m_\tau) \, 
\nn\\
&
\>\times  
(\cos\De + \I \ga_5 \sin\De) \, 
\nn\\
&
\>\times  
(-\sla{p}_{\tau^+\!} + m_\tau) \,
(\sla{p}_{\pi^+\!} - \sla{p}_{\pi^{0+\!}}) \, 
P_\LL v_{\nu^+}
\,,\eql{FullM:initial}
\end{align}
where $\pi^{0\pm}$ refers to the $\pi^0$ coming from the $\rho^\pm$
decay, respectively, and we have denoted $\nu_\tau$ and
$\bar{\nu}_\tau$ as $\nu^-$ and $\nu^+$, respectively.  The following
approximations have been made above:
\begin{itemize} 
\item{We neglected the diagram in which the two $\pi^0$ are exchanged,
  assuming that we can identify $\pi^{0\pm}$ by looking for a $\pi^0$
  flying near $\pi^\pm$, respectively.  As the taus from $h \to \tau^+
  \tau^-$ are highly boosted and back-to-back in the Higgs rest frame,
  this should be an excellent approximation.}
\item{All intermediate particles are assumed to be on-shell, so the
  denominators of their propagators have been dropped
  in~\eq{FullM:initial}, as they are just momentum-independent
  constants $\sim \I m \Ga$.}
\item{We neglect $m_{\pi^\pm} - m_{\pi^0}$ throughout the paper.  A
  convenient consequence of this (very good) approximation is that the
  $\rho \pi \pi$ amplitude in~\eq{M:rho_to_pi-pi} effectively
  satisfies a ``Ward identity'', {\it i.e.}, it vanishes upon
  replacing $\ep_{\rho^-}$ with $p_{\rho^-}$:
\begin{align}
p_{\rho^-} \dt (p_{\pi^-\!} - p_{\pi^{0-\!}})
=
m_{\pi^\pm}^2 - m_{\pi^0}^2
=
0
\,.\eql{WI}
\end{align}
This is why we have dropped the $p_\mu p_\nu / m_\rho^2$ term of the
$\rho^\pm$ propagators in~\eq{FullM:initial}.}
\end{itemize}
Carefully keeping the combinations $p_{\pi^\pm\!} - p_{\pi^{0\pm}}$ intact
as suggested by the heuristic analysis of \Sec{heuristic}, the
amplitude~\eq{FullM:initial} can be rewritten as
\begin{align}
\cM_\text{full}
\propto\>\>
\bar{u}_{\nu^-} \,
\sla{q}_-
\bigl( \e^{\I \De} \sla{p}_{\tau^-} - \e^{-\I \De} \sla{p}_{\tau^+\!} \bigr)
\sla{q}_+ 
P_\LL v_{\nu^+}
\,,\eql{FullM:final}
\end{align}
where
\begin{align}
q_\pm \equiv p_{\pi^\pm} - p_{\pi^{0\pm}}
\,.\eql{var:q+-}
\end{align}

Taking $\{ p_{\tau^\pm},\, q_\pm,\, p_{\nu^\pm} \}$ as the set of
independent variables (subject to the constraint $(p_{\tau^+\!} +
p_{\tau^-})^2 = m_h^2$), let us analyze how the physics depends on
these momenta.  First, in the square of the
amplitude~\eq{FullM:final}, the variables $q_\pm$ and $p_{\nu^\pm}$
will only enter via the products $\sla{q}_+ \sla{p}_{\nu^+} \sla{q}_+$
and $\sla{q}_- \sla{p}_{\nu^-} \sla{q}_-$.  These combinations can be
further simplified as
\begin{align}
\sla{q}_\pm \sla{p}_{\nu^\pm} \sla{q}_\pm
=
(m_\tau^2 + m_\rho^2)\, \sla{k}_\pm
\,,
\end{align}
where 
\begin{align}
k_\pm^\mu
\equiv 
y_\pm \, q_\pm^\mu
+ r \, p_{\nu^\pm}^\mu
\eql{var:k+-}
\end{align}
with%
\footnote{$y_{+,-}$ are respectively equal to $y_{1,2}$ used in
  Refs.~\cite{Bower:2002zx, Worek:2003zp, Desch:2003mw,
    Desch:2003rw}.}
\begin{align}
y_\pm
&\equiv 
\fr{2 q_\pm \dt p_{\tau^\pm}}{m_\tau^2 + m_\rho^2}
=
\fr{q_\pm \dt p_{\tau^\pm}}{p_{\rho^\pm} \dt p_{\tau^\pm}}
\,,\eql{var:y+-}\\%%
r 
&\equiv 
\fr{m_\rho^2 - 4 m_\pi^2}{m_\tau^2 + m_\rho^2}
\approx 
0.14
\,.\eql{var:r}
\end{align}
In terms of $k_\pm$ and $p_{\tau^\pm}$, the square of the amplitude
in~\eq{FullM:final} only involves the traces over four $\gamma$
matrices, and an elementary computation gives
\begin{align}
|\cM|^2
\propto\>\>
\Pboring + \Puseless + \Pgood + \Pgood^*
\,,\eql{matrix_element:full}
\end{align}
where
\begin{align}
\Pboring
&\equiv 
2\bigl[ 
(k_- \dt p_{\tau^-}) (p_{\tau^-} \dt k_+)
+ (k_+ \dt p_{\tau^+}) (p_{\tau^+} \dt k_-)
\nn\\
&\hspace{4.6ex}
- m_\tau^2 \, (k_- \dt k_+)
\bigr]
\,,\\%%
\Puseless
&\equiv
-2 \cos(2\De) \, 
(k_- \dt p_{\tau^-}) (k_+ \dt p_{\tau^+})
\,,\\%%
\Pgood 
&\equiv
- \e^{2\I \De}
\bigl[
 (k_- \dt p_{\tau^+}) (k_+ \dt p_{\tau^-}) 
-(p_{\tau^-} \dt p_{\tau^+}) (k_- \dt k_+)
\nn\\
&\hspace{9.3ex}
-\I \ep_{\mu\nu\rho\sg} \, k_-^\mu \, p_{\tau^-}^\nu k_+^\rho \, p_{\tau^+}^\sg
\bigr].
\eql{Pgood:original}
\end{align}
Here, $\Pgood$ is the interesting contribution that depends on both
$\De$ and the $\tau^\pm$ spins.  On the other hand, $\Pboring$ is an
uninteresting piece since it is independent of $\De$. (It is sensitive
to the $\tau^\pm$ spins, {\it i.e.}, the relative orientation of the
$\tau^+$ and $\tau^-$ subsystems, as it involves scalar products like
$\dsp{k_- \dt k_+}$).  Lastly, $\Puseless$ does depend on $\De$ but is
insensitive to the spins, as it only involves $\dsp{k_+ \dt
  p_{\tau^+}}$ and $\dsp{k_- \dt p_{\tau^-}}$, which are just scalar
quantities of the $\tau^+$ and $\tau^-$ subsystems alone.

We therefore focus on $\Pgood$.  To reveal how it depends on the
relative orientations of the $\tau^\pm$ systems to each other, observe
that $\Pgood$ is antisymmetric under $k_\pm \leftrightarrow
p_{\tau^\pm}$.  This suggests that $k_\pm$ and $p_{\tau^\pm}$ should
be combined into two antisymmetric tensors $F_\pm^{\mu\nu}$, one for
each $\tau^\pm$ system:
\begin{align}
F_\pm^{\mu\nu}
\equiv
k_\pm^\mu \, p_{\tau^\pm}^\nu - k_\pm^\nu \, p_{\tau^\pm}^\mu
=
- F_\pm^{\nu\mu}
\,.
\end{align}
In terms of these, $\Pgood$ takes an elegant form:
\begin{align}
\Pgood
=
\e^{2\I \De}
\Bigl( 
  \fr12 F_{-\mu\nu} F_+^{\mu\nu} 
+ \fr{\I}{4} \ep_{\mu\nu\rho\sg} \, F_-^{\mu\nu} F_+^{\rho\sg}
\Bigr).
\end{align}
Moreover, the fact that $F_\pm^{\mu\nu}$ are antisymmetric 2nd-rank
tensors suggests that the physics is clearest in terms of their
``electric'' and ``magnetic'' components:
\begin{align}
E_\pm^i 
\equiv 
F_\pm^{i0}
\,,\quad
B_\pm^i 
\equiv 
-\fr12 \ep^{ijk} F_{\pm jk}
\,.\eql{E_and_B}
\end{align}
Indeed, $\Pgood$ then simplifies into just one term:
\begin{align}
\Pgood
&=
-\e^{2\I \De}
\bigl[ 
(\vec{E}_- + \I\vec{B}_-) \dt (\vec{E}_+ + \I\vec{B}_+)
\bigr].
\eql{Pgood}
\end{align}

We will now develop intuition for $\vec{E}_\pm$ and $\vec{B}_\pm$.
First, from~\eq{E_and_B}, we have
\begin{align}
\vec{B}_\pm
=
\vec{p}_{\tau^\pm} \!\times \vec{k}_\pm 
=
\vec{v}_{\tau^\pm} \!\times \vec{E}_\pm 
\,,
\eql{B}
\end{align}
where $\vec{v}_{\tau^\pm} \equiv \vec{p}_{\tau^\pm} / p_{\tau^\pm}^0$
is the 3-velocity of the $\tau^\pm$.  Thus, $\vec{B}_\pm =0$ in the
rest frame of each $\tau^\pm$, respectively, while in all other frames
$\vec{B}_\pm$ are perpendicular to both $\vec{E}_\pm$ and
$\vec{p}_{\tau^\pm}$.  Moreover, in the boosted $\tau^\pm$ limit
($|\vec{v}_{\tau^\pm}| \to 1$), we have $|\vec{B}_\pm| =
|\vec{E}_\pm|$.

Second, from~\eq{E_and_B}, $\vec{E}_\pm$ is given by
\begin{align}
\vec{E}_\pm
&=
p_{\tau^\pm}^0 \, \vec{k}_\pm  - k_\pm^0 \, \vec{p}_{\tau^\pm}
\,.\eql{E}
\end{align}
Clearly, $\vec{E}_\pm$ takes the simplest form in the $\tau^\pm$ rest
frame since then $\vec{p}_{\tau^\pm}$ in the second term vanishes.
Let us use $\rest$ to indicate the quantities evaluated in the
respective $\tau^\pm$ rest frames.  Then, combining~\eq{var:k+-}
and~\eq{E} in the $\tau^\pm$ rest frames, we have
\begin{align}
\vec{E}_\pm\rest 
&=
m_\tau \, \vec{k}_\pm\rest 
\nn\\%%
&=
m_\tau 
\Bigl[ 
(y_\pm - r) \, \vec{p}_{\pi^\pm}\rest 
- (y_\pm + r) \, \vec{p}_{\pi^{0\pm}}\rest 
\Bigr] 
\,,\eql{E_rest}
\end{align}
where we have used $\vec{p}_{\nu^\pm}\rest = -\vec{p}_{\pi^\pm}\rest -
\vec{p}_{\pi^{0\pm}}\rest$.  Therefore, in an arbitrary frame with a
$\tau^\pm$ velocity $\vec{v}_{\tau^\pm}$, we have
\begin{align}
\vec{E}_{\pm}^{||}
&=
\vec{E}_{\pm}^{||}\rest
\,,\nn\\
\vec{E}_{\pm}^\perp
&=
\gamma_\pm  
\Bigl[ 
\vec{E}_{\pm}\rest - \vec{v}_{\tau^\pm} \!\times\! \vec{B}_\pm\rest 
\Bigr]^\perp
=
\fr{E_{\tau^\pm}}{m_\tau} \, \vec{E}_{\pm}^\perp\rest
\,,\eql{E_components} 
\end{align}
where $\vec{E}_\pm^{||}$ and $\vec{E}_\pm^\perp$ are the components of
$\vec{E}_\pm$ parallel and perpendicular to $\vec{v}_{\tau^\pm}$,
respectively, while $\gamma_\pm \equiv (1 -
|\vec{v}_{\tau^\pm}|^2)^{-1/2} = E_{\tau^\pm}/m_\tau$.  An important
implication of~\eq{E_components} is that, for a boosted $\tau^\pm$
($E_{\tau^\pm} / m_\tau \gg 1$), we get $|\vec{E}_\pm^\perp| \gg
|\vec{E}_\pm^{||}|$, so $\vec{E}_\pm$ also becomes perpendicular to
$\vec{v}_{\tau^\pm}$.  Thus, the relative magnitudes and orientations
of $\vec{E}_\pm$, $\vec{B}_\pm$, and $\vec{v}_{\tau^\pm}$ in the
boosted $\tau^\pm$ limit are akin to those of electromagnetic waves.

To summarize, we write out $\vec{E}_\pm$ and $\vec{B}_\pm$ in the
Higgs rest frame.  Since the $\tau^\pm$ are highly boosted in this
frame, we can neglect $E_\pm^{||}$.  Then, combining~\eq{E_rest}
and~\eq{E_components} with $E_{\tau^\pm} = m_h /2$, we get
\begin{align} 
\vec{E}_\pm
&=
\fr{m_h}{2} 
\Bigl[ 
(y_\pm - r) \, \vec{p}_{\pi^\pm}\rest 
- (y_\pm + r) \, \vec{p}_{\pi^{0\pm}}\rest 
\Bigr]^\perp 
\,,\eql{E:higgs_frame}
\end{align}
where $\rest$ on a vector indicates that the vector should be
evaluated in the respective $\tau^\pm$ rest frame, while $^\perp$
denotes the components perpendicular to the respective $\tau^\pm$
velocity in the Higgs rest frame.  Recall $\vec{B}_\pm$ is given 
by~\eq{B}. 

\begin{center}
\begin{figure}
\includegraphics[width=0.9\linewidth]{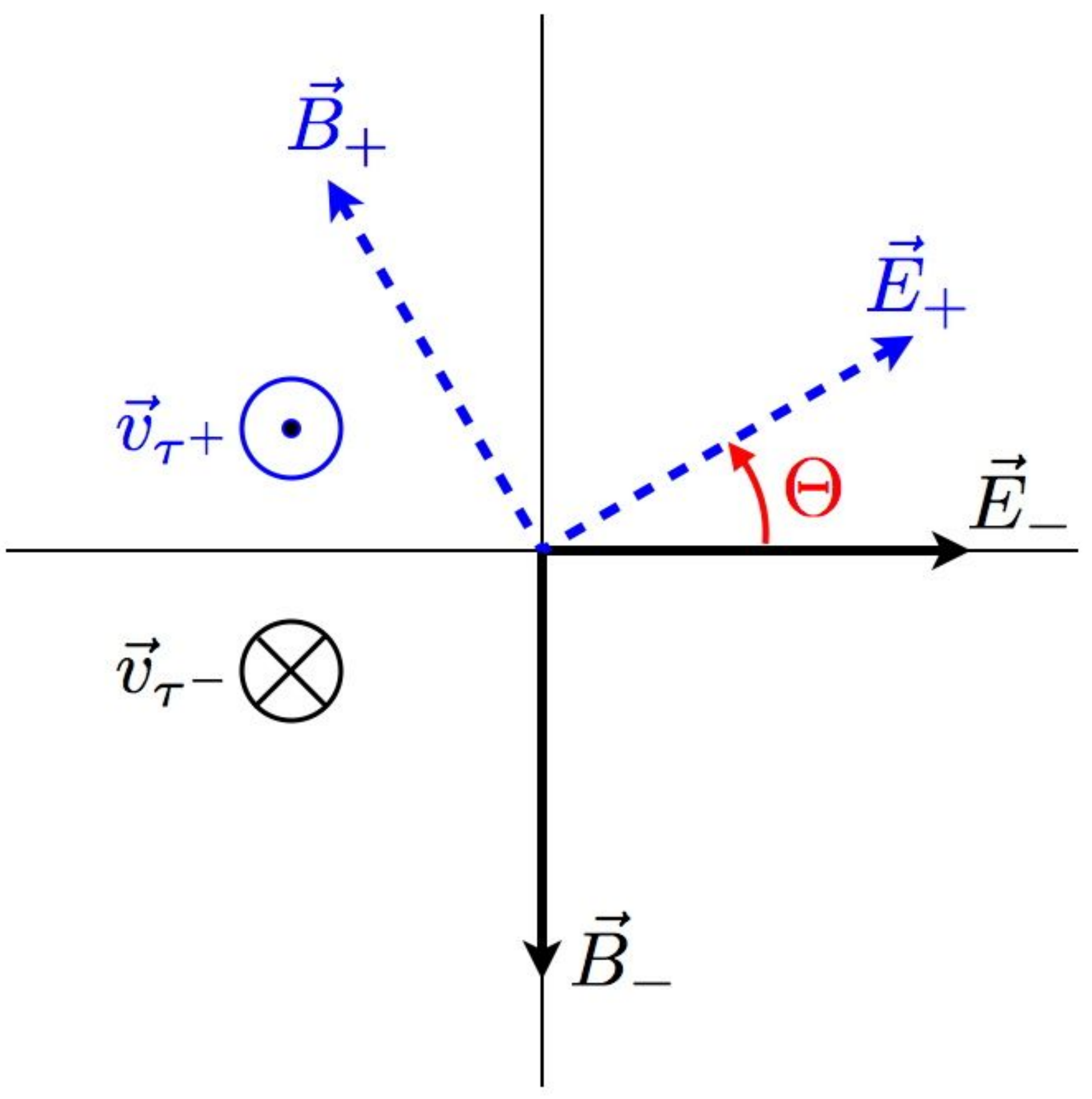}
\caption{\Figl{Theta_definition} The definition of our variable
  $\Tha$, drawn in the Higgs rest frame with the $\tau^-$ and $\tau^+$ going in and out of the page respectively. Note $\Tha$ is taken to be
  positive if $\vec{E}_+$ is on the upper-half plane, and negative
  otherwise, where $\vec{E}_-$ is fixed to lie along the $+\hat{x}$
  axis.}
\end{figure}
\end{center}
%

%%%%%%%%%%%%%%%%%%%%%%%%%%%%%%%%%%%%%%%%%%%%%%%%%%%%%%%%%%%%%%%%%%%%%%%%%%%%%%%
%%%%%%%%%%%%%%%%%%%%%%%%%%%%%%%%%%%%%%%%%%%%%%%%%%%%%%%%%%%%%%%%%%%%%%%%%%%%%%%
\subsection{The $\Tha$ angle}
\Secl{Theta}
%%%%%%%%%%%%%%%%%%%%%%%%%%%%%%%%%%%%%%%%%%%%%%%%%%%%%%%%%%%%%%%%%%%%%%%%%%%%%%%
%%%%%%%%%%%%%%%%%%%%%%%%%%%%%%%%%%%%%%%%%%%%%%%%%%%%%%%%%%%%%%%%%%%%%%%%%%%%%%%
We are ready to evaluate $\Pgood$ in the Higgs rest frame.  In this
frame, since $\vec{v}_{\tau^+}$ and $\vec{v}_{\tau^-}$ are back to
back, the $\vec{E}_+$--$\vec{B}_+$ plane and the
$\vec{E}_-$--$\vec{B}_-$ plane are parallel to each other.  Thus, we
will superimpose them to make a single plane.  In this combined plane,
let $\vec{E}_-$ and $\vec{B}_-$ point to the right and downward,
respectively. (See \Fig{Theta_definition}.)  Then, we define $\Tha$ to
be the angle of $\vec{E}_+$ with respect to $\vec{E}_-$, where $0 \leq
\Tha \leq \pi$ if $\vec{E}_+$ is on the upper-half plane, while $-\pi
< \Tha < 0$ if on the lower-half plane.%
\footnote{In other words, $\Tha$ is 
the acoplanarity angle between the $\vec{E}_+$--$\vec{v}_{\tau^+}$ plane 
and $\vec{E}_-$--$\vec{v}_{\tau^-}$ plane, where the orientation of 
the planes defined by the respective $\vec{B}_\pm$.}
That is,
\begin{align}
\Tha
=
\sgn\Bigl[ \vec{v}_{\tau^+} \dt (\vec{E}_- \!\times\! \vec{E}_+) \Bigr]
\Arccos\biggl[ 
\fr{\vec{E}_+ \dt \vec{E}_-}
   {\Abs{\vec{E}_{+\!}} \, \Abs{\vec{E}_{-\!}}} 
\biggr],\eql{Thetadef}
\end{align}
where $\Arccos$ takes values between $0$ and $+\pi$.  
Then $\vec{B}_+$
makes an angle $\Tha + \pi/2$ with respect to $\vec{E}_-$.  The
magnitudes of $\vec{B}_\pm$ are the same as the respective
$\vec{E}_\pm$.  Putting everything together, the
distribution~\eq{Pgood} becomes
\begin{align}
\Pgood
=
-2 \e^{\I(2\De - \Tha)} \, \Abs{\vec{E}_{+\!}} \, \Abs{\vec{E}_{-\!}} 
\,,\eql{GreatFormula}
\end{align}
where $\vec{E}_\pm$ are given by~\eq{E:higgs_frame}.  The
contributions that have been neglected to arrive at~\eq{GreatFormula}
from~\eq{Pgood:original} are only $\cO(m_\tau^2 / m_h^2) \sim 0.01$\%.

%%%%%%%%%%%%%%%%%%%%%%%%%%%%%%%%%%%%%%%%%%%%%%%%%%%%%%%%%%%%%%%%%%%%%%%%%%%%%%%
%%%%%%%%%%%%%%%%%%%%%%%%%%%%%%%%%%%%%%%%%%%%%%%%%%%%%%%%%%%%%%%%%%%%%%%%%%%%%%%
%%%%%%%%%%%%%%%%%%%%%%%%%%%%%%%%%%%%%%%%%%%%%%%%%%%%%%%%%%%%%%%%%%%%%%%%%%%%%%%
\section{Collider Studies}
\Secl{studies}
%%%%%%%%%%%%%%%%%%%%%%%%%%%%%%%%%%%%%%%%%%%%%%%%%%%%%%%%%%%%%%%%%%%%%%%%%%%%%%%
%%%%%%%%%%%%%%%%%%%%%%%%%%%%%%%%%%%%%%%%%%%%%%%%%%%%%%%%%%%%%%%%%%%%%%%%%%%%%%%
%%%%%%%%%%%%%%%%%%%%%%%%%%%%%%%%%%%%%%%%%%%%%%%%%%%%%%%%%%%%%%%%%%%%%%%%%%%%%%%

In this section we develop collider analyses aimed at reconstructing
the $\Theta$ angle in~\eq{Thetadef}.  From~\eq{matrix_element:full}
and~\eq{GreatFormula}, the matrix element squared for the $h \to \pi^+
\pi^0 \bar{\nu} \pi^- \pi^0 \nu$ decay has a term proportional to
$-\cos (\Theta - 2\Delta)$: the $\Theta$ distribution is thus
sensitive to the $CP$ phase $\Delta$ as its minimum is located at $2
\Delta$.  As before, we fix $y_\tau \equiv y_\tau^{\text{SM}}$ and
therefore the only new parameter we introduce is $\Delta$.

We implement the $\Delta$ phase in~\eq{Lpheno} and the effective
vertices in~\eq{M:tau_to_rho-nu} and~\eq{M:rho_to_pi-pi} into a
FeynRules v.1.6.0~\cite{Christensen:2008py} model.  We then generate
Monte Carlo events in MadGraph 5~\cite{Alwall:2011uj} for $p\, p \to
h\, j$ production at the LHC with $\sqrt{s} = 14$ TeV as well as $e^+
\, e^- \to Z\, h$ production at the ILC with $\sqrt{s} = 250$ GeV: in
either case, the Higgs decays via $h \to \pi^+ \pi^0 \bar{\nu} \pi^-
\pi^0 \nu$.  In order to retain quantum interference effects, the full
$2\to 7$ body process is simulated.  For the LHC study, we also
generate a background sample of $p\, p \to Z\, j$ production with the
subsequent decay $Z \to \pi^+ \pi^0 \bar{\nu} \pi^- \pi^0 \nu$.

We will first study the effectiveness of the $\Theta$ distribution at
truth level, assuming the neutrino momenta are known: this facilitates
a comparison to the $\phi^*$ variable~\cite{Bower:2002zx,
  Worek:2003zp}, which was previously proposed for studying $CP$
violation in the Higgs coupling to taus.  After demonstrating the
superior qualities of the $\Theta$ variable, we present a sensitivity
study for reconstructing $\Theta$ at the ILC, where the neutrino
four-momentum can be reconstructed up to a two-fold ambiguity.
Finally, we turn to the LHC, where the neutrinos cannot be
reconstructed and the irreducible $Z$ background is significant.  In
this case, we find that using a collinear
approximation~\cite{Ellis:1987xu} for the neutrino momenta in addition
to the standard hard cuts for Higgs events still allows the $\Theta$
distribution to retain significant discrimination power between
different underlying $\Delta$ signal models.

We do not include pileup or perform any detector simulation in this
work, aside from implementing flat efficiencies for $\tau$-tagging for
the LHC study.  Pileup effects are expected to complicate the primary
vertex determination necessary for measuring charged pion tracks as
well as contribute extra ambient radiation in the electromagnetic
calorimeter (ECAL), making neutral pion momenta measurements more
difficult.  Furthermore, finite tracking and calorimeter resolutions
are expected to smear the $\Theta$ distribution.  In particular, the
ability to distinguish between charged and neutral pion momenta when
both pions are overlapping also could affect the $\Theta$ measurement.
Note, however, that because of the magnetic field, the softer
$\pi^\pm$ and $\pi^0$ could be separated at the ECAL\@.  Even if the
two pions overlap in the ECAL, the $\pi^0$ momentum can be obtained by
subtracting the track momentum from the total momentum measured in
ECAL, assuming negligible contamination from other sources of energy
deposition.

We also neglect the neutral pion combinatoric issue, which is
justified if the respective parent rho mesons are boosted far apart as
a result of the Higgs decay.  In general, the $\pi^\pm$ and $\pi^0$
coming from the same $\rho^\pm$ parent are mostly collinear.  This
fact has been exploited in the hadronic tau tagging algorithm.  For
example, the HPS algorithm used by CMS requires that the charged and
neutral hadrons are contained in a cone of the size $\Delta R = (2.8
\text{ GeV}/c)/\pT^{\tau_h}$, where $\pT^{\tau_h}$ is the transverse
momentum of the reconstructed tau~\cite{Chatrchyan:2012zz}.  Since the
two tau candidates are usually required to be well separated, the
combinatorics problem in determining the correct $\rho^\pm$ parents
can be ignored.

%%%%%%%%%%%%%%%%%%%%%%%%%%%%%%%%%%%%%%%%%%%%%%%%%%%%%%%%%%%%%%%%%%%%%%%%%%%%%%%
%%%%%%%%%%%%%%%%%%%%%%%%%%%%%%%%%%%%%%%%%%%%%%%%%%%%%%%%%%%%%%%%%%%%%%%%%%%%%%%
\subsection{Truth level}
%%%%%%%%%%%%%%%%%%%%%%%%%%%%%%%%%%%%%%%%%%%%%%%%%%%%%%%%%%%%%%%%%%%%%%%%%%%%%%%
%%%%%%%%%%%%%%%%%%%%%%%%%%%%%%%%%%%%%%%%%%%%%%%%%%%%%%%%%%%%%%%%%%%%%%%%%%%%%%%

Recall from~\eq{matrix_element:full} and~\eq{GreatFormula} that the
minimum of the $\Theta$ distribution is located at $2 \Delta$, and so
constructing the $\Theta$ distribution allows us to read off the
$\Delta$ phase of the underlying signal model.  In
\Fig{True_varyDelta}, we show the $\Theta$ distribution in $p\, p \to
h\, j$ events where we have temporarily assumed the neutrinos are
fully reconstructed.  The various signal models with $\Delta = 0$
($CP$-even), $\Delta = \pi / 4$ (maximal $CP$ admixture), and $\Delta
= \pi /2$ ($CP$-odd) clearly show the large $-\cos(\Theta - 2 \Delta)$
contribution of the matrix element as seen in~\eq{GreatFormula}.  We
also superimpose the $\Theta$ distribution from $p\, p \to Z\, j$
event. Note that it is flat.  Clearly, observing the cosine
oscillation in experimental data will require both a favorable signal
to background ratio as well as a solution for the neutrino momenta
that preserves the inherently large amplitude of the $\Theta$
oscillation.
\begin{center}
\begin{figure}
\includegraphics[width=0.9\linewidth]{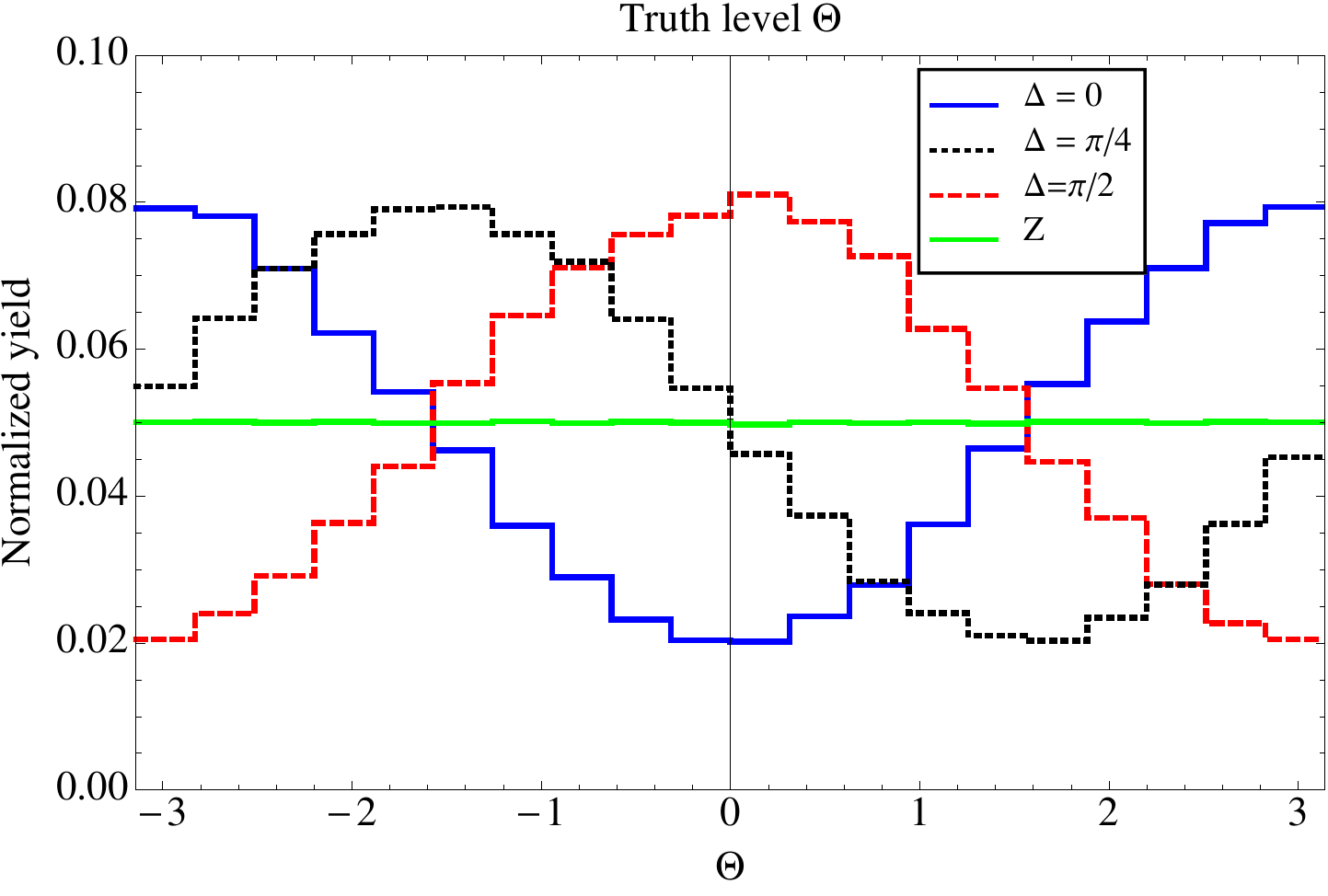}
\caption{\Figl{True_varyDelta} The $\Theta$ distributions (compare
  with~\eq{GreatFormula}) for the Higgs with $\Delta = 0$ ($CP$-even),
  $\Delta = \pi / 4$ (maximal $CP$ admixture), and $\Delta = \pi / 2$
  ($CP$-odd), and the $Z$, assuming neutrinos are fully reconstructed.
  The relative normalization of the $Z$ line is arbitrary.}
\end{figure}
\end{center}
We now compare $\Theta$ at truth level with the $\phi^*$ variable
proposed in Refs.~\cite{Bower:2002zx, Worek:2003zp}: here, $\phi^*$ is
the acoplanarity angle between the decay planes of $\rho^+$ and
$\rho^-$ in the $\rho^+ \rho^-$ rest frame.  The sign of $\phi^*$ is
defined as the sign of the product of $\vec p_{\pi^-} \cdot (\vec
p_{\pi^+} \times \vec p_{\pi^0} )$.  Following~\cite{Bower:2002zx,
  Worek:2003zp}, the events are divided into two classes, $y_+ y_- <
0$ and $y_+ y_- > 0$, where the two classes are differ by a
$180^\circ$ phase shift.  In order to make a direct comparison with
our $\Theta$ variable, we combine the $\phi^*$ distributions of the
two classes with a $180^\circ$ phase shift so the phases of the two
classes agree.  Note that while $\phi^*$ does not refer to the
neutrinos, this classification into the two classes still requires the
knowledge of the neutrino momenta (see \eq{var:y+-}).  Assuming the
neutrinos are fully reconstructed, the $\Theta$ and $\phi^*$
distributions for $p\, p \to h\, j$ events are shown in
\Fig{ThetavsWorek} with $\Delta = 0$.  We readily see that oscillation
amplitude of the $\Theta$ distribution is larger than that of the
acoplanarity angle $\phi^*$ by about $50\%$.  Compared to $\phi^*$,
the $\Theta$ variable thus provides superior sensitivity to the $CP$
phase $\Delta$.
\begin{center}
\begin{figure}
\includegraphics[width=0.9\linewidth]{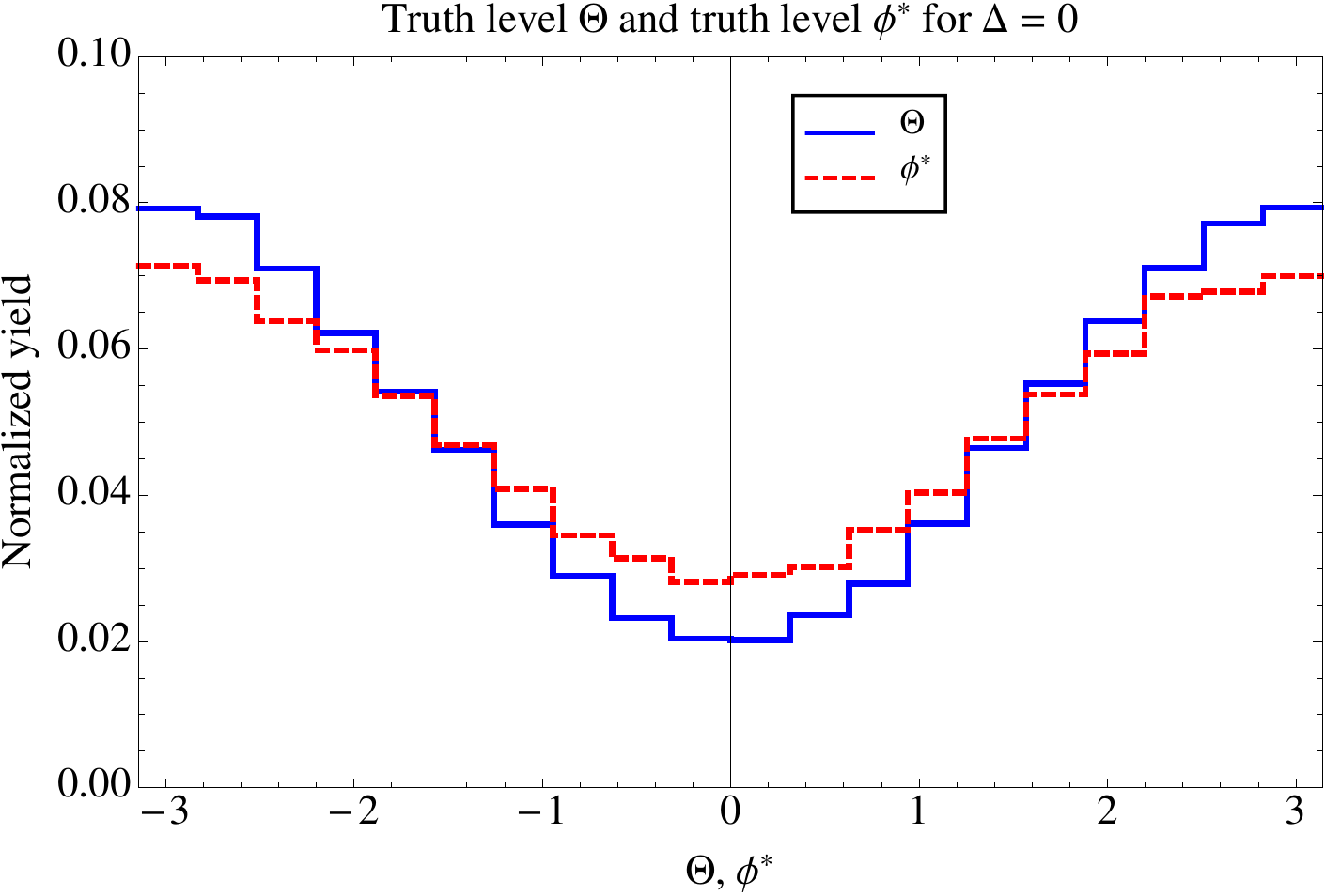}
\caption{\Figl{ThetavsWorek} The distributions of our $\Theta$ and the
  $\phi^*$ variable of Ref.~\cite{Bower:2002zx, Worek:2003zp} for
  $\Delta = 0$.  The $\phi^*$ distribution is aggregated from the two
  $y_+ y_- > 0$ and $y_+ y_- < 0$ classes as explained in the text to
  make the direct comparison clearer.}
\end{figure}
\end{center}
Having considered the case where the neutrinos from the tau decays are
fully reconstructed, we next turn to the lepton collider environment,
where we will find the neutrinos can be fully reconstructed up to a
two-fold ambiguity.

%%%%%%%%%%%%%%%%%%%%%%%%%%%%%%%%%%%%%%%%%%%%%%%%%%%%%%%%%%%%%%%%%%%%%%%%%%%%%%%
%%%%%%%%%%%%%%%%%%%%%%%%%%%%%%%%%%%%%%%%%%%%%%%%%%%%%%%%%%%%%%%%%%%%%%%%%%%%%%%
\subsection{An $e^+e^-$ Higgs Factory}
%%%%%%%%%%%%%%%%%%%%%%%%%%%%%%%%%%%%%%%%%%%%%%%%%%%%%%%%%%%%%%%%%%%%%%%%%%%%%%%
%%%%%%%%%%%%%%%%%%%%%%%%%%%%%%%%%%%%%%%%%%%%%%%%%%%%%%%%%%%%%%%%%%%%%%%%%%%%%%%

At a lepton collider running at $\sqrt{s} = 250$ GeV, such as the ILC,
the main production mode for the Higgs is via associated production
with a $Z$ boson.  Our prescribed decay mode for the Higgs, $h \to
\pi^+ \, \pi^0 \,\bar{\nu}\, \pi^- \, \pi^0 \,\nu$, has two neutrinos
that escape the detector.  We use the known initial four momenta, two
tau mass and two neutrino mass constraints to solve for each neutrino
momentum component.  Note we will assume the $Z$ decays to visible
states, which will reduce our event yield by 20\%.  Solving the system
of equations for the neutrino momenta gives rise to a two-fold
ambiguity, where one solution is equal to the truth input neutrino
momenta while the other gives a set of wrong neutrino momenta.  Note
both solutions are consistent with four-momentum conservation and
therefore correctly reconstruct the Higgs mass.  Since these solutions
are indistinguishable in the analysis, we assign each solution half an
event weight.

The resulting distribution of $\Theta$ for $\Delta = 0$ is given in
\Fig{ILCTheta}, where we superimpose the truth level $\Theta$
distribution for $e^+ e^- \to Z h$ events for easy comparison.  We can
see that the oscillation amplitude at the ILC is degraded from the
truth level result by $\sim 30\%$.  We also show the reconstructed
distribution for $\Delta = 0$, $\Delta = \pi / 4$, and $\Delta =
\pi/2$ in \Fig{ILCvaryDelta}.  While the two-fold ambiguity for the
neutrino momenta solution set does degrade the truth level result, the
{\it reconstructable} $\Theta$ distribution in \Fig{ILCvaryDelta}
shows significant discrimination power between various $\Delta$ signal
models.  Note the amplitude of pseudoscalar distribution ($\Delta =
\pi/2$) is slightly higher than the scalar amplitude: here, the
``wrong solution'' approximates the correct neutrino momenta on
average better than the other $\Delta = 0$ or $\Delta = \pi / 4$
cases. This small effect can be traced back to
equation~\eq{M:h_to_tautau:higgs_frame} where we derived that a
pseudoscalar decays to two taus in the singlet spin state. As a
result, in this case the two tau spins point in opposite directions,
regardless of the spin quantization axis. In the pseudoscalar case the
two tau decays thus tend to occur with opposite orientation and the
two neutrinos are slightly more back-to-back and consequently the two
solutions for their momenta are closer together.

\begin{center}
\begin{figure}
\includegraphics[width=0.9\linewidth]{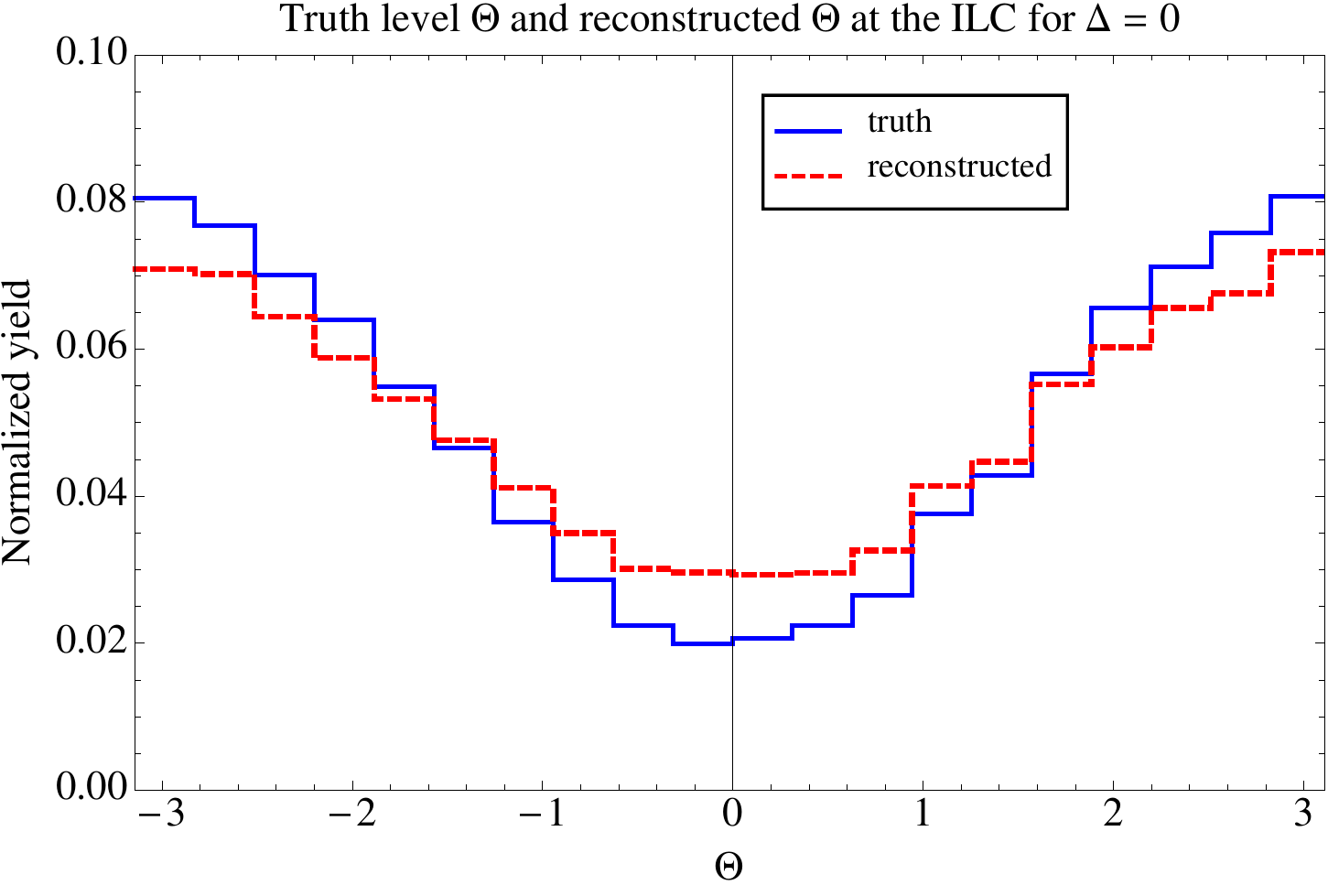}
\caption{\Figl{ILCTheta} The truth and reconstructed $\Theta$
  distributions at the ILC for $\Delta = 0$.}
\end{figure}
\end{center}

\begin{center}
\begin{figure}
\includegraphics[width=0.9\linewidth]{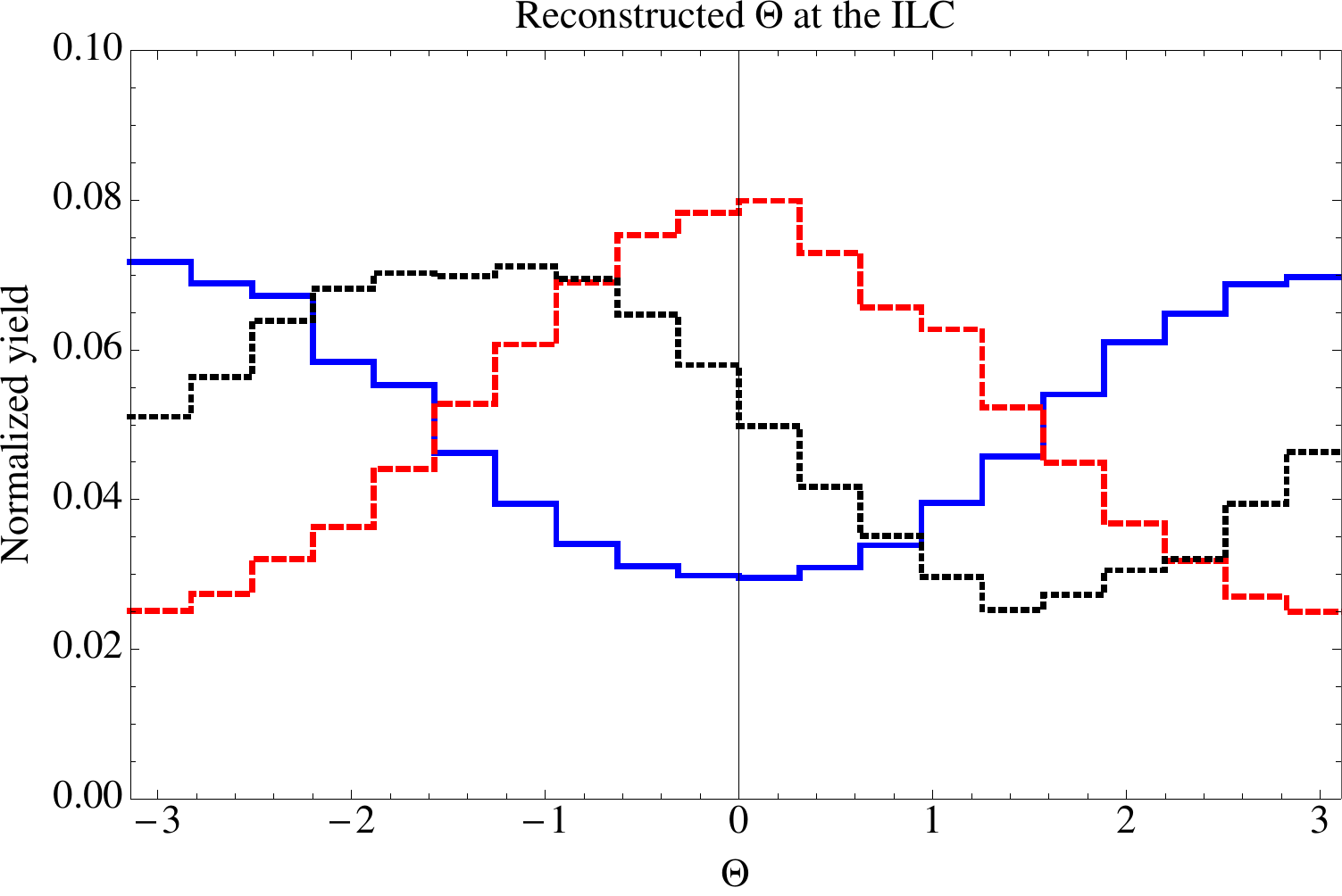}
\caption{\Figl{ILCvaryDelta} The reconstructed $\Theta$ distribution
  at the ILC for $\Delta = 0$, $\Delta = \pi/4$, and $\Delta =
  \pi/2$.}
\end{figure}
\end{center}

We now discuss the projected ILC sensitivity for measuring $\Delta$.
At the ILC, the cross section for $Z h$ production at $\sqrt{s} = 250$
GeV with polarized beams $P(e^-, e^+) = (-0.8, 0.3)$ for $m_h = 125$
GeV is 0.30 pb~\cite{ILCTDR}.%
\footnote{We have checked the $\Theta$ distribution is insensitive to
  the polarization of the $e^-$-$e^+$ beams.}
Assuming a Higgs branching fraction to tau pairs of 6.1\%, a $\tau^-
\to \rho^- \nu \to \pi^- \pi^0 \nu$ branching fraction of 26\%, and a
$Z$-to-visible branching fraction of 80\%, we calculate the ILC should
have 990 events with 1 ab$^{-1}$ of luminosity.  Since the solved
neutrino momenta correctly reconstruct the Higgs mass, the $ZZ$
backgrounds are negligible and will be ignored.

\begin{table}
   \centering
   %\topcaption{Table captions are better up top} % requires the topcapt package
   \parbox{4.5cm}{
    \begin{ruledtabular}
   \begin{tabular}{cc} % Column formatting, @{} suppresses leading/trailing space
 $\sigma_{e^+ e^- \to h Z}$ & 0.30 pb \\%298.2 fb \\
 Br($h \rightarrow \tau^+ \tau^-$) & 6.1\% \\
 Br($\tau^- \rightarrow \pi^- \pi^0 \nu$) & 26\% \\
 Br($Z \to $ visibles) & 80\% \\
\hline
 N$_{\text{events}}$ & 990 \\
\hline
 Accuracy & $4.4^\circ$ \\
   \end{tabular}
      \end{ruledtabular}}
   \caption{Cross section, branching fractions, expected number of
     signal events, and accuracy for measuring $\Delta$ for the ILC
     with $\sqrt{s} = 250$ GeV and 1 ab$^{-1}$ integrated
     luminosity.}
   \Tabl{ILC}
\end{table}

To estimate the expected ILC accuracy for measuring $\Delta$, we
perform a log likelihood ratio test for the SM hypothesis with $\Delta
= 0$ against an alternative hypothesis with $\Delta = \delta$.  In
general, the likelihood ratio in $N$ bins is given by
\begin{align}
L
&=
\fr{ \prod\limits_{i=1}^{N} \operatorname{Pois} 
\left( B_i + S_i^{\Delta =0} | B_i + S_i^{\Delta = \delta} \right) }{ 
\prod\limits_{i=1}^{N}  \operatorname{Pois} 
\left( B_i + S_i^{\Delta =0} | B_i + S_i^{\Delta =0} \right) } 
\,,\eql{LR}
\end{align}
where $B_i$, $S_i^{\Delta = 0}$ and $S_i^{\Delta = \delta}$ are the
number of background events, signal events assuming $\Delta = 0$, and
signal events assuming $\Delta = \delta$ in bin $i$ of the $\Theta$
distribution.  In our ILC treatment, we neglect $ZZ$ and $Z \to \tau
\tau$ continuum backgrounds and so we set $B_i = 0$.  Here,
$\operatorname{Pois} (k|\lambda)$ is the usual Poisson distribution
function, $\operatorname{Pois} (k|\lambda) = \lambda^k e^{-\lambda} /
k!$.

We parametrize the signal $\Theta$ distribution with
a~$\displaystyle{c-A \cos(\Theta - 2 \Delta)}$ fit function, where the
offset constant $c$ and oscillation amplitude $A$ are fixed by the fit
of the standard model $\Theta$ distribution with $\Delta = 0$, giving
$c_0$ and $A_0$ respectively.  Then, the resulting $S^{\Delta =
  \delta}$ signal $\Theta$ distribution is given by $c_0 - A_0
\cos(\Theta - 2 \delta)$.  We construct the binned
likelihood\footnote{We choose $N=100$ bins, though we verified the
  number of bins is immaterial for our results.} according to~\eq{LR}
for various $\delta$ hypotheses to test the discrimination against the
SM hypothesis.  With 1 ab$^{-1}$ of ILC luminosity, we find $1 \sigma$
discrimination at $\delta = 0.077 \text{ rad} = 4.4^\circ$, which is a
highly promising degree of sensitivity for measuring the $CP$ phase of
the Higgs coupling to taus.  We summarize our rate estimate and
accuracy result in \Tab{ILC}.

We remark that this sensitivity estimate is only driven by statistical
uncertainties, and systematic uncertainties are expected to reduce the
efficacy of our result.  Also, detector resolution effects and SM
backgrounds, while expected to be small, will also slightly degrade
our projection.  Based on our results, which surpass earlier accuracy
estimates of $6^\circ$~\cite{Desch:2003rw}, a full experimental
sensitivity study incorporating these subleading effects is certainly
warranted.

%%%%%%%%%%%%%%%%%%%%%%%%%%%%%%%%%%%%%%%%%%%%%%%%%%%%%%%%%%%%%%%%%%%%%%%%%%%%%%%
%%%%%%%%%%%%%%%%%%%%%%%%%%%%%%%%%%%%%%%%%%%%%%%%%%%%%%%%%%%%%%%%%%%%%%%%%%%%%%%
\subsection{LHC}
%%%%%%%%%%%%%%%%%%%%%%%%%%%%%%%%%%%%%%%%%%%%%%%%%%%%%%%%%%%%%%%%%%%%%%%%%%%%%%%
%%%%%%%%%%%%%%%%%%%%%%%%%%%%%%%%%%%%%%%%%%%%%%%%%%%%%%%%%%%%%%%%%%%%%%%%%%%%%%%

We now develop an LHC study for reconstructing the $\Theta$
distribution in $p\, p \to h\, j$ in the $\pi^+ \pi^0 \pi^- \pi^0 + j
+ \MET$ final state. We use the $h+j$ final state for a couple of
reasons. First, since hadronic taus can be faked by jets, $pp \to h
\to $ two hadronic taus faces an immense background from multijet
QCD. By requiring another object in the final state, we gain handles
to suppress the background.  Second, the collinear approximation gives
ambiguous results if the two taus are back-to-back, so the requirement
of an additional object in the event guarantees we are away from this
configuration.  One option is associated production of a Higgs wit a
$W/Z$. However this rate is quite small, especially once the branching
ratios for $W/Z$ into clean final states are taken into account.
Other possibilities include Higgs production via vector boson fusion
and in association with a jet. Both of these options give promising
signal-to-background ratios and both should be considered.  For
concreteness we will consider $pp \to h+j$ here as a demonstration of
the feasibility of our technique.

As mentioned before, the neutrinos are not reconstructible in the
hadron collider environment, and so we will employ the collinear
approximation~\cite{Ellis:1987xu} for the neutrino momenta.  In
\Fig{Truthvscollinear}, we show a comparison between the truth level
$\Theta$ distribution and the $\Theta$ distribution using the
collinear approximation for neutrino momenta, for the $\Delta = 0$
benchmark.  While the collinear approximation reduces the oscillation
amplitude of the distribution, the location of the minimum of the
distribution does not change.  Therefore, measuring $\Delta$ is a
viable possibility at the LHC using the collinear approximation for
the neutrino momenta.  We remark that in the collinear approximation,
$\Theta$ is equivalent to the acoplanarity angle
$\phi^*$~\cite{Bower:2002zx, Worek:2003zp}.  Yet, we are the first
feasibility study for measuring $CP$ violation in the Higgs coupling
to taus at hadron colliders using prompt tau decays and kinematics.
With a more sophisticated scheme than the collinear approximation, the
$\Theta$ variable will be superior to $\phi^*$.

\begin{center}
\begin{figure}
\includegraphics[width=0.9\linewidth]{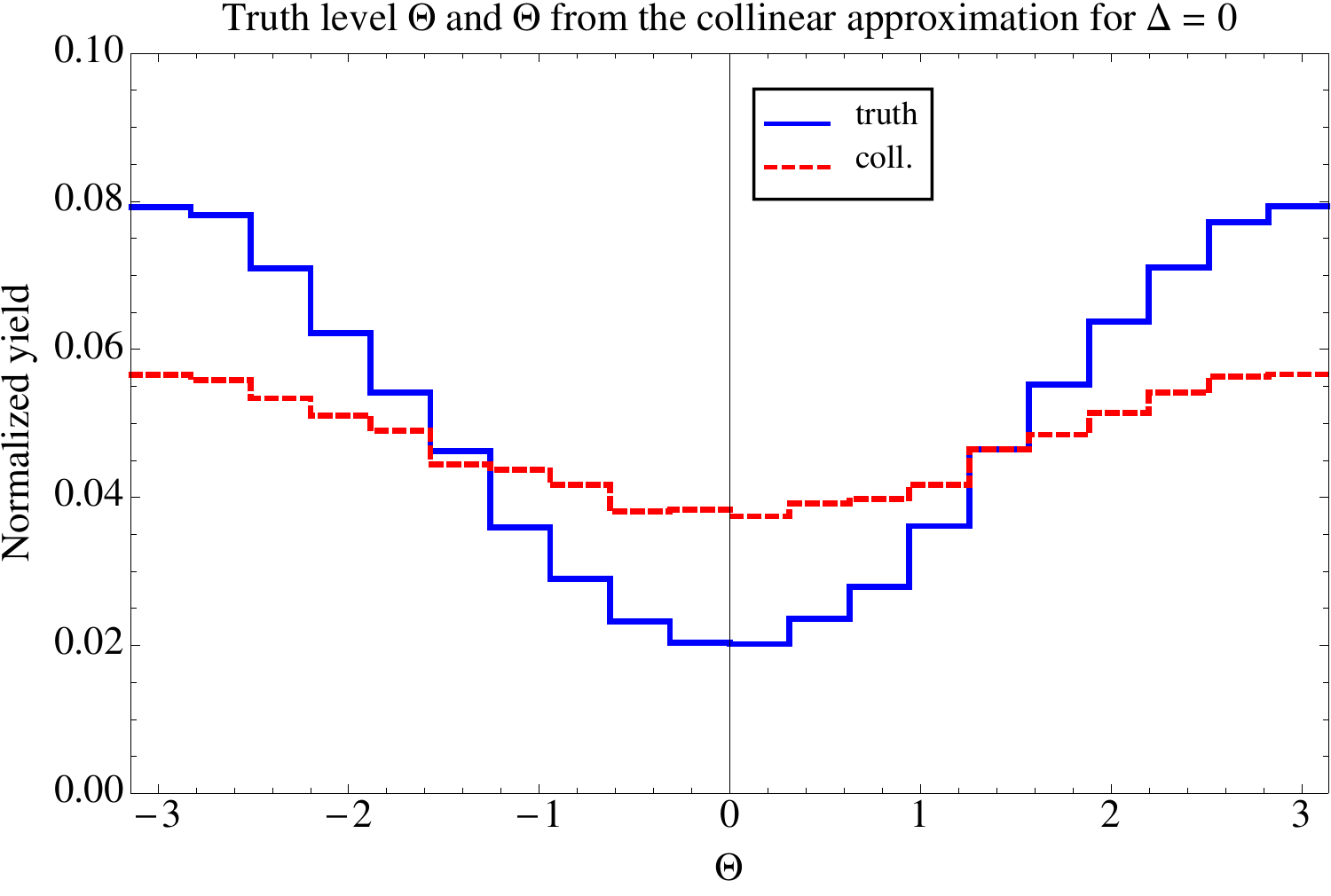}
\caption{\Figl{Truthvscollinear} The distributions of the
  truth-$\Theta$ and $\Theta$ from the collinear approximation for
  $\Delta = 0$.}
\end{figure}
\end{center}

At the LHC, the dominant background for the $h\, j$ signal process is
the irreducible $Z\, j$ background, where the $Z$ decays to the same
final state as the higgs.  As shown earlier in \Fig{True_varyDelta},
the $\Theta$ distribution from $Z$ events is flat: importantly, this
is true regardless of possible mass window cuts on the reconstructed
$m_{\tau \tau}$ resonance.  We remark that the $CP$ phase in the Higgs
coupling to taus does manifest in the $Z$--$\tau$--$\tau$ vertex at
one loop.  Since this effect is suppressed by $\sim
y_{\tau}^2/(16\pi^2) \sim \mathcal O(10^{-4})$, whereas the signal to
background ratio will be $\mathcal O (60\%)$, we can safely ignore the
loop induced $CP$ phase in the $Z$--$\tau$--$\tau$ vertex.  In
addition, we will assume that the QCD background contribution also has
a flat $\Theta$ distribution, since the QCD contamination in the
signal region is not expected to have any particular spin
correlations.

Using our $h\, j$ and $Z\, j$ event samples from MadGraph 5 for a 14
TeV LHC, we first isolate the signal region with a series of hard
cuts.  First, we apply a preselection requirement on the leading jet
$\pT > 140$ GeV with $|\eta| < 2.5$.  Using MCFM
v.6.6~\cite{Campbell:2011bn} with these preselection requirements on
the leading jet, we obtain a $h\, j$ NLO inclusive cross section of
2.0 pb with $m_h = 126$ GeV and a $Z\, j$ NLO inclusive cross section
of 420 pb.  After applying the appropriate Higgs, $Z$, and tau
branching fractions, we calculate a signal cross section of 8.2 fb and
$Z$ background cross section of 970 fb.%
\footnote{These numbers were generated using CTEQ6M parton
  distribution functions. For the signal we use a
  factorization/renormalization scale of $\mu_F = m_H/2$, while for
  the background we use $\mu_F = \sqrt{M^2_Z + p^2_{\mathrm{T},j}
  }$. These scale choices are motivated by agreement with higher order
  (NNLO) calculations (where they exist).}
Next, we impose hard kinematic cuts to isolate the signal.  Motivated
by~\cite{CMS:utj}, we choose the signal region to be:
\begin{itemize}
  \item $\MET > 40$ GeV,
  \item $\pT^{\rho^\pm} > 45$ GeV,
  \item $|\eta^{\rho^\pm}| < 2.1$,
  \item $m_{\text{coll}} > 120$ GeV,
\end{itemize}
where $m_{\text{coll}}$ is the reconstructed Higgs mass by using the
collinear approximation. The hard $m_{\text{coll}}$ cut strongly
suppresses the $Z+j$ background, but is less effective on multijet
QCD\@. To reduce the multijet component -- and its accompanying
uncertainty -- to less than 10\% of the total background we impose a
high $\MET$ cut.  The net efficiencies for signal and $Z$ background
after these cuts are 18\% and 0.24\%, respectively. Rather than
simulate the QCD contribution, we account for QCD contamination in the
signal region by increasing the $Z$ background rate by 10\%: a
complete treatment of the expected QCD background is beyond the scope
of this study.  Finally, for hadronic $\tau$ tagging efficiency, we
consider a standard 50\% efficiency and a more optimistic 70\%
efficiency~\cite{Chatrchyan:2012zz}.  We therefore expect 1100 signal
events and 1800 $Z + $ QCD background events with 3 ab$^{-1}$ of
luminosity from the 14 TeV LHC, assuming 50\% $\tau$ tagging
efficiency.  These rates are summarized in \Tab{input}.

%%%%%%%%%%%%%%
\begin{table}
   \centering
   %\topcaption{Table captions are better up top} % requires the topcapt package
   \parbox{7cm}{
     \begin{ruledtabular}
   \begin{tabular}{ccc} % Column formatting, @{} suppresses leading/trailing space
          & $h\, j$ & $Z\, j$ \\
\hline
 Inclusive $\sigma$ & 2.0 pb & 420 pb \\
 Br($\tau^+ \tau^-$ decay) & 6.1\% & 3.4\% \\
 Br($\tau^- \to \pi^- \pi^0 \nu$) & 26\% & 26\% \\
 Cut efficiency & 18\% & 0.24\% \\
\hline
 N$_{\text{events}}$ & 1100 & 1800 \\
   \end{tabular}
   \end{ruledtabular}}
   \caption{Cross sections, branching fractions, cut efficiencies, and
     expected number of events assuming 3 ab$^{-1}$ and 50\% $\tau$
     tagging efficiency for the Higgs signal and the $Z$ background:
     the background number of events includes an additional 10\%
     contribution from QCD multijet background.}  \Tabl{input}
\end{table}

We note that although we generated signal and background samples
independently, there is a small interference between Higgs and $Z$
diagrams in the $g q \to \tau^+ \tau^- q$ diagram.  Our checks of this
interference on the $\Theta$ distributions for combined signal and
background events versus separate signal and background events showed
a negligible effect: we thus ignore this interference effect.

We now perform a likelihood analysis~\eq{LR} to quantify how
effectively the $\Theta$ distribution distinguishes between signal
hypotheses with different $CP$ phases in the presence of $Z + $ QCD
background.  First, we test the discrimination between a pure scalar
and a pure pseudoscalar $h$--$\tau$--$\tau$ coupling.  We find that
these two hypotheses can be distinguished at $3 \sigma$ sensitivity
with 550 (300) fb$^{-1}$ assuming 50\% (70\%) $\tau$ tagging
efficiency.  We can attain $5 \sigma$ sensitivity between pure scalar
and pseudoscalar couplings with 1500 (700) fb$^{-1}$ luminosity
assuming 50\% (70\%) efficiency.

We also estimate the possible accuracy for the LHC experiments to
measure $\Delta$ with an upgraded luminosity of $3\iab$.  We adopt the
same procedure as with the ILC accuracy estimate described in the
previous section, modified to account for the $Z + $ QCD background,
which is fixed to be flat in $\Theta$.  We find that the accuracy in
measuring $\Delta$ is $11.5^\circ$ ($8.0^\circ$) assuming 50\% (70\%)
hadronic $\tau$ tagging efficiency.  The scalar versus pseudoscalar
discrimination and the accuracy estimates are summarized in \Tab{LHC}.

%%%%%%%%%%%%%%
\begin{table}
   \centering
   %\topcaption{Table captions are better up top} % requires the topcapt package
      \parbox{8.5cm}{
   \begin{ruledtabular}
   \begin{tabular}{ccc} % Column formatting, @{} suppresses leading/trailing space
   $\tau_{h}$ efficiency & 50\% & 70\% \\
   \hline
   $3 \sigma$ &  $L = 550 \text{ fb}^{-1}$  & $L = 300 \text{ fb}^{-1}$ \\
   $5 \sigma$ &  $L = 1500 \text{ fb}^{-1}$  & $L = 700 \text{ fb}^{-1}$ \\
      \hline
   Accuracy($L=3\iab$) &  $11.5^\circ$  & $8.0^\circ$ \\
   \end{tabular}
   \end{ruledtabular}}
   \caption{The luminosity required for distinguishing the scalar and
     pseudoscalar couplings and the accuracy in measuring $\Delta$
     with $3\iab$ of luminosity at the 14 TeV LHC\@.}
   \Tabl{LHC}
\end{table}

Again, these estimates are based only on statistical uncertainties
without performing a full detector simulation.  The effects from
pileup and detector resolution are expected to degrade these
projections, but corresponding improvements in the analysis, such as a
more precise approximation for the neutrino momenta, improved
background understanding (from other LHC measurements) or multivariate
techniques, could counterbalance the decrease in sensitivity.  The
promising results of our study strongly motivate a comprehensive
analysis by the LHC experiments for the prospect of measuring the $CP$
phase~$\Delta$.

%%%%%%%%%%%%%%%%%%%%%%%%%%%%%%%%%%%%%%%%%%%%%%%%%%%%%%%%%%%%%%%%%%%%%%%%%%%%%%%
%%%%%%%%%%%%%%%%%%%%%%%%%%%%%%%%%%%%%%%%%%%%%%%%%%%%%%%%%%%%%%%%%%%%%%%%%%%%%%%
%%%%%%%%%%%%%%%%%%%%%%%%%%%%%%%%%%%%%%%%%%%%%%%%%%%%%%%%%%%%%%%%%%%%%%%%%%%%%%%
\section{Conclusions}
\Secl{conclusion}
%%%%%%%%%%%%%%%%%%%%%%%%%%%%%%%%%%%%%%%%%%%%%%%%%%%%%%%%%%%%%%%%%%%%%%%%%%%%%%%
%%%%%%%%%%%%%%%%%%%%%%%%%%%%%%%%%%%%%%%%%%%%%%%%%%%%%%%%%%%%%%%%%%%%%%%%%%%%%%%
%%%%%%%%%%%%%%%%%%%%%%%%%%%%%%%%%%%%%%%%%%%%%%%%%%%%%%%%%%%%%%%%%%%%%%%%%%%%%%%

Higgs decays to tau leptons provide a singular opportunity to measure
the $CP$ properties of the Higgs-fermion couplings.  In this paper, we
have studied the decay of $h \rightarrow \tau^+\tau^-$ followed by
$\tau^\pm \rightarrow (\rho^{\pm} \rightarrow \pi^\pm \pi^0)\, \nu$. A
new observable, $\Theta$, was constructed in~\eq{Thetadef} using the
momenta of the tau decay products.  The differential cross section can
be written in a form of $c - A \cos(\Theta - 2\Delta)$, hence the
$\Theta$ distribution can be used to distinguish various $CP$ mixing
as shown in~\Fig{True_varyDelta}.  The $\Theta$ variable can be viewed
as an acoplanarity angle between the planes spanned by certain linear
combinations of the pion and neutrino momenta, and it was demonstrated
to be superior to previously proposed acoplanarity angles.

At the ILC, where the neutrino momenta can be reconstructed up to a
two-fold ambiguity, the advantages of the $\Theta$ variable are most
evident.  We estimate that the $CP$ phase can be measured to an
accuracy of $4.4^\circ$ for $\sqrt{s} = 250$ GeV\@, a substantial
improvement over previous results.  For the LHC, we have had to rely
on the collinear approximation to reconstruct the neutrino momenta and
some of the discriminating power of the $\Theta$ variable is lost.
Nevertheless, we find an accuracy of $11.5^\circ$ ($8.0^\circ$) is
possible after 3000 fb$^{-1}$ of luminosity and assuming a
50\%\,(70\%) tau tagging efficiency.  Recasting in terms of the
parameters introduced in~\eq{Leff}, a $5^\circ$--$10^\circ$ deviation
from the SM case is equivalent to sensitivity to $\Lambda \sim 10$
TeV, where $\Lambda$ is the scale of the dimension six operator
in~\eq{Leff} and $|\beta|$ is assumed to be $\mathcal{O}(1)$.  A
better approximation scheme for the neutrino momenta will improve
these results.

In our collider studies we have neglected detector effects and
background systematic uncertainties. While adding these effects will
worsen our results, this may be offset by better understanding of the
backgrounds (thereby allowing looser cuts) and with a more
sophisticated ({\it e.g.} MVA) analysis and statistical tools.

Finally, in this work we picked specific Higgs production mechanisms
and focused on a single decay channel.  To understand the full extent
of future colliders' sensitivities to the $CP$ phase of Higgs-fermion
couplings, additional production channels such as VBF should be
explored, both at the LHC and in a Higgs factory.  In addition, other
hadronic decay channels, as well as semi-leptonic channels, of the tau
pair might also be sensitive to the $CP$ properties of the Higgs.

%%%%%%%%%%%%%%%%%%%%%%%%%%%%%%%%%%%%%%%%%%%%%%%%%%%%%%%%%%%%%%%%%%%%%%%%%%%%%%%
%%%%%%%%%%%%%%%%%%%%%%%%%%%%%%%%%%%%%%%%%%%%%%%%%%%%%%%%%%%%%%%%%%%%%%%%%%%%%%%
%%%%%%%%%%%%%%%%%%%%%%%%%%%%%%%%%%%%%%%%%%%%%%%%%%%%%%%%%%%%%%%%%%%%%%%%%%%%%%%
\acknowledgments
%%%%%%%%%%%%%%%%%%%%%%%%%%%%%%%%%%%%%%%%%%%%%%%%%%%%%%%%%%%%%%%%%%%%%%%%%%%%%%%
%%%%%%%%%%%%%%%%%%%%%%%%%%%%%%%%%%%%%%%%%%%%%%%%%%%%%%%%%%%%%%%%%%%%%%%%%%%%%%%
%%%%%%%%%%%%%%%%%%%%%%%%%%%%%%%%%%%%%%%%%%%%%%%%%%%%%%%%%%%%%%%%%%%%%%%%%%%%%%%
We would like to thank Kaustubh Agashe, Wolfgang Altmannshofer, Yuval
Grossman, Uli Haisch, Josh Ruderman, Daniel Stolarski, Raman Sundrum,
Ciaran Williams and Jure Zupan for comments and discussions.  We would
also like to thank the Kavli Institute for Theoretical Physics at UCSB
where part of this work was performed.  RP is supported by the NSF
under grant PHY-0910467.  TO is supported by the DOE under grant
DE-FG02-13ER41942.  Fermilab is operated by the Fermi Research
Alliance, LLC under Contract No.~De-AC02-07CH11359 with the United
States Department of Energy.

%%%%%%%%%%%%%%%%%%%%%%%%%%%%%%%%%%%%%%%%%%%%%%%%%%%%%%%%%%%%%%%%%%%%%%%%%%%%%%%
%%%%%%%%%%%%%%%%%%%%%%%%%%%%%%%%%%%%%%%%%%%%%%%%%%%%%%%%%%%%%%%%%%%%%%%%%%%%%%%
%%%%%%%%%%%%%%%%%%%%%%%%%%%%%%%%%%%%%%%%%%%%%%%%%%%%%%%%%%%%%%%%%%%%%%%%%%%%%%%
\appendix
\section{A Simple UV Completion of the Dimension-6 Operator}
\Appl{UVcompletion}
%%%%%%%%%%%%%%%%%%%%%%%%%%%%%%%%%%%%%%%%%%%%%%%%%%%%%%%%%%%%%%%%%%%%%%%%%%%%%%%
%%%%%%%%%%%%%%%%%%%%%%%%%%%%%%%%%%%%%%%%%%%%%%%%%%%%%%%%%%%%%%%%%%%%%%%%%%%%%%%
%%%%%%%%%%%%%%%%%%%%%%%%%%%%%%%%%%%%%%%%%%%%%%%%%%%%%%%%%%%%%%%%%%%%%%%%%%%%%%%

In this appendix we give an example for a UV completion for the
dimension-6 operator in~\eq{Leff}, {\it i.e.}, the term with $\be$.
Our purpose is not to advocate a specific model as particularly
well-motivated but to simply provide an existence proof of a
weakly-coupled renormalizable theory that can generate the $\be$ term
in~\eq{Leff} at $\La \sim 1 \TeV$ with an arbitrary $CP$ phase,
without generating other operators that may contradict with
experiments.

Consider an extension of the SM with a second higgs doublet $\Phi$
with $m_\Phi \gsim 1\TeV$ with the following tree-level lagrangian:
\begin{align}
\cL_\text{tree}
=&\;
\cL_{\text{SM}-y_\tau}
\nn\\
&+|\DD\Phi|^2 - m_\Phi^2 |\Phi|^2 - \la_\Phi |\Phi|^4
\eql{UVcomp}\\
&-(y H\ell_{3\LL}^\dag \tau_\RR^\PD
+y' \Phi \ell_{3\LL}^\dag \tau_\RR^\PD
+\la' (\Phi^{\dag\!} H) |H|^2
+\cc)
\,,\nn
\end{align}
where $\cL_{\text{SM}-y_\tau}$ is the SM lagrangian without the tau
Yukawa coupling.  The full quantum lagrangian is $\cL_\text{tree} +
\cL_\text{ct}$, where $\cL_\text{ct}$ contains all counterterms
\emph{necessary} for consistent renormalization at loop level.  For
simplicity, we neglect neutrino masses and mixings, so
$\cL_\text{tree}$ possesses an accidental $\gu{1}_e \times \gu{1}_\mu
\times \gu{1}_\tau$ family symmetry, which is then inherited by
$\cL_\text{ct}$ as well.  This immediately implies that there are no
lepton flavor changing processes such as $\mu \to e\ga$.  There are no
constraints from quark flavor/$CP$ measurements; since the couplings
of $\Phi$ to quarks are absent in $\cL_\text{tree}$ and only appear in
$\cL_\text{ct}$, they are not only very small ($\sim y'/16\pi^2$
$\times$ the corresponding SM Yukawa) but also respect the CKM flavor
structure of the SM\@.  Similarly, the couplings of $\Phi$ to $e$ and
$\mu$ are inconsequential; in particular, we have checked that a
contribution to the electron elecric dipole moment induced at 2-loop
level is negligible.  Finally, the modification of the coupling of $Z$
to $\tau$ is also tiny, safely below the LEP constraints.

In order to see the effects of~\eq{UVcomp} on Higgs decays let us
consider the limit in which $m_\Phi\gg v$ and the doublet $\Phi$ can
be integrated out and we can consider an effective field theory below
$m_\Phi$.  At tree level, this generates two dimension-6 interactions:
\begin{align}
\cL_\text{dim-6} 
=
\fr{|\la'|^2}{m_\Phi^2} |H|^6 
+ 
\Bigl( \fr{\la' y'}{m_\Phi^2} |H|^2 H\ell_{3\LL}^\dag \tau_\RR^\PD + \cc \Bigr).
\end{align}
This theory now matches on to the effective theory~\eq{Leff} with $\al
= y$, $\be = y'\la'$ and $\La = m_\Phi$.  It should be noted that this
theory contains in general a $CP$ violating phase. In particular, the
phase of $y^*y'\lambda'$ may not be rotated away by field
redefinitions.  Taking $\La \sim \TeV$ and $|\la'| \sim |y'| \sim 1$
with arbitrary phases in $y$, $y'$ and $\la'$ can therefore produce an
$\cO(1)$ $CP$-violating phase in Higgs decays to tau leptons.

Other theories, including composite Higgs models~\cite{Agashe:2004rs,
  Giudice:2007fh} and models with vector-like
leptons~\cite{Kearney:2012zi} may also produce the necessary
interactions to induce $CP$ violating Higgs decays into taus. In
constructing such models one should take care that the coupling of $Z$
to taus is not modified above the $10^{-3}$ level (see
Ref.~\cite{ALEPH:2005ab} or Fig.~10.4 of~\cite{Beringer:1900zz})
either by construction (as in the model we just discussed) or by a
cancellation among various contributions.

%%%%%%%%%%%%%%%%%%%%%%%%%%%%%%%%%%%%%%%%%%%%%%%%%%%%%%%%%%%%%%%%%%%%%%%%%%%%%%%
%%%%%%%%%%%%%%%%%%%%%%%%%%%%%%%%%%%%%%%%%%%%%%%%%%%%%%%%%%%%%%%%%%%%%%%%%%%%%%%
%%%%%%%%%%%%%%%%%%%%%%%%%%%%%%%%%%%%%%%%%%%%%%%%%%%%%%%%%%%%%%%%%%%%%%%%%%%%%%%
%\bibliographystyle{apsrev}
%\bibliography{bibliography}{99}

\end{document}